\documentclass[authoryear,review]{elsarticle}
\usepackage{appendix}
\usepackage{setspace}
\usepackage[font=small,  margin=2cm]{caption}
\usepackage[tbtags]{amsmath}
\usepackage{amsthm,amssymb,amsfonts}
\usepackage{natbib}
\usepackage{multirow}
\usepackage{lscape}
\journal{Arxiv}
\usepackage{subfigure}
\usepackage{makecell}
\usepackage{exscale}
\usepackage{booktabs}
\usepackage{array}
\usepackage{fullpage}
\usepackage{url}
\usepackage{algorithm}
\usepackage{algpseudocode}
\usepackage{bm}
\usepackage{smile}
\usepackage{mathtools}
\usepackage{wrapfig}
\usepackage{lipsum}
\usepackage{mathrsfs}
\usepackage{relsize}
\usepackage{dsfont}
\usepackage{multirow}
\usepackage{apalike}
\usepackage{chngcntr}
\usepackage{graphicx}
\usepackage{algorithmicx}
\usepackage{algpseudocode}
\usepackage[usenames,dvipsnames,svgnames,table]{xcolor}
\usepackage[colorlinks=true,
linkcolor=blue,
urlcolor=blue,
citecolor=blue]{hyperref}

\numberwithin{equation}{section}
\numberwithin{theorem}{section}
\numberwithin{corollary}{section}
\counterwithout{asmp}{section}
\numberwithin{definition}{section}

\begin{document}
	
	\begin{frontmatter}
		\title{  Robust Statistical Inference for Large-dimensional Matrix-valued Time Series via Iterative Huber Regression}

		\author[myfirstaddress]{Yong He\corref{cor1}}
		\address[myfirstaddress]{Institute for Financial Studies, Shandong University, Jinan, 250100, China}
			\cortext[cor1]{Corresponding author. All authors contributed equally to this work.}
		\ead{heyong@sdu.edu.cn}
	\author[myfourthaddress]{Xin-Bing Kong}
		\address[myfourthaddress]{Nanjing Audit University, Nanjing, 211815, China}
	
		\ead{xinbingkong@126.com}
	\author[mysecondaddress]{Dong Liu}
		\address[mysecondaddress]{Shanghai University of Finance and
Economics,  Shanghai, 200433, China}
			\author[mythirdaddress]{Ran Zhao}
		\address[mythirdaddress]{School of Mathematics, Shandong University, Jinan, 250100, China}


  \begin{abstract}
   Matrix factor model is drawing growing attention for simultaneous two-way dimension reduction of well-structured matrix-valued observations. This paper focuses on robust statistical inference for matrix factor model in the ``diverging dimension" regime. We derive the convergence rates of the robust estimators for loadings, factors and common components  under finite second moment assumption of the idiosyncratic errors. In addition, the asymptotic distributions of the estimators are also derived under mild conditions. We propose a rank minimization and an eigenvalue-ratio method to estimate the pair of factor numbers consistently.
   Numerical studies confirm the iterative Huber regression algorithm is a practical and reliable approach for the estimation of matrix factor model, especially under the cases with heavy-tailed idiosyncratic errors . We illustrate the  practical  usefulness of the proposed methods by two real datasets, one on financial portfolios and one on the macroeconomic indices of China. 
 	
	\end{abstract}
	\end{frontmatter}

	\section{Introduction}
Modern multivariate (vector) time series typically consist of a large number of variables and dimension reduction is of great importance for extracting useful information from these large datasets.
 Factor model is an extremely powerful tool of summarizing information from large datasets and large-dimensional (approximate) factor model draws growing attention in the ``big-data" era since the seminal work by \cite{bai2002determining} and \cite{stock2002forecasting}.  Henceforth, there is a flourishing  trend on this topic  during the last two decades, including but not limited to \cite{bai2003inferential}, \cite{onatski09}, \cite{ahn2013eigenvalue}, \cite{fan2013large}, \cite{Trapani2018A}, \cite{Barigozzi2018Simultaneous}, \cite{Sahalia2017Using}, \cite{Sahalia2020High}, \cite{Barigozzi2020Sequential} and \cite{fan2022learning}. The aforementioned works typically assume that the fourth moment (or even higher moment) of the factors and idiosyncratic errors is finite, which may be restrictive in real applications of finance and economics. To the best of our knowledge,
 there exist a few works on relaxing the moment conditions, see for example, the endeavors by \cite{yu2019robust}, \cite{Chen2021Quantile} and \cite{He2020large}.

In contrast to multivariate vector time series, matrix time series are growing common in various research areas and can be defined as a sequence of $p_1\times p_2$ random matrices $\{\Xb_t, 1\leq t\leq T\}$ with each random matrix
used to model observations that are well structured to be an array. An example in macroeconomics is
the import-export volumes between countries. Another example is a series of
macroeconomic indicators (GDP, inflation, interest rates...) for multiple countries. In marketing studies, the recommender system is based on customers' ratings of satisfaction on a large number of items, as time elapses, resulting in a series of huge rating matrices. In
finance, portfolio returns are sorted in size levels and book-to-equity ratio levels. The past few years have seen increasing interest in large-dimensional matrix factor models, since the seminal work by \cite{wang2019factor},
who proposed a two-way factor structure for matrix-valued time series. \citet{wang2019factor} proposed estimators of the factor loading matrices and numbers of the row and column factors based on an eigen-analysis of the auto-cross-covariance matrix, extending the theoretical analysis framework of \cite{LY12} to the matrix factor setting.
\citet{fan2021} proposed an $\alpha$-PCA method by conducting eigen-analysis of a weighted average of the sample mean and the column (row) sample covariance matrix; \citet{Yu2021Projected} further proposed a projected estimation method and improved the estimation efficiency of the factor loading matrices.  \cite{He2021MatrixFA} provide a least square interpretation of the PE method by \cite{Yu2021Projected}, which parallels to the least-square interpretation of the PCA for the vector factor model. They further extend the
least squares to minimizing the Huber loss function, and proposed a weighted iterative projection approach
to compute and learn the parameters (RMFA hereafter), which is robust to heavy-tailed idiosyncratic errors. For other
 extensions and applications of the matrix factor model, one may refer to \cite{Chen2020Modeling},  \citet{chen2019constrained},  \citet{liu2019helping}, \cite{Gao2021A}, \cite{Jing2021Community}. Recently, a growing number of papers on the broad context of tensor factor model come to appear, see for example, \cite{han2020rank,chen2020semiparametric,Han2020Tensor,han2021cp,lam2021rank,tensorTS,zhang2022tucker,chenlam2022,chen2022factor,Chang2021Modelling}.

\begin{figure}[!h]
    \centerline{\includegraphics[width=14cm,height=10cm]{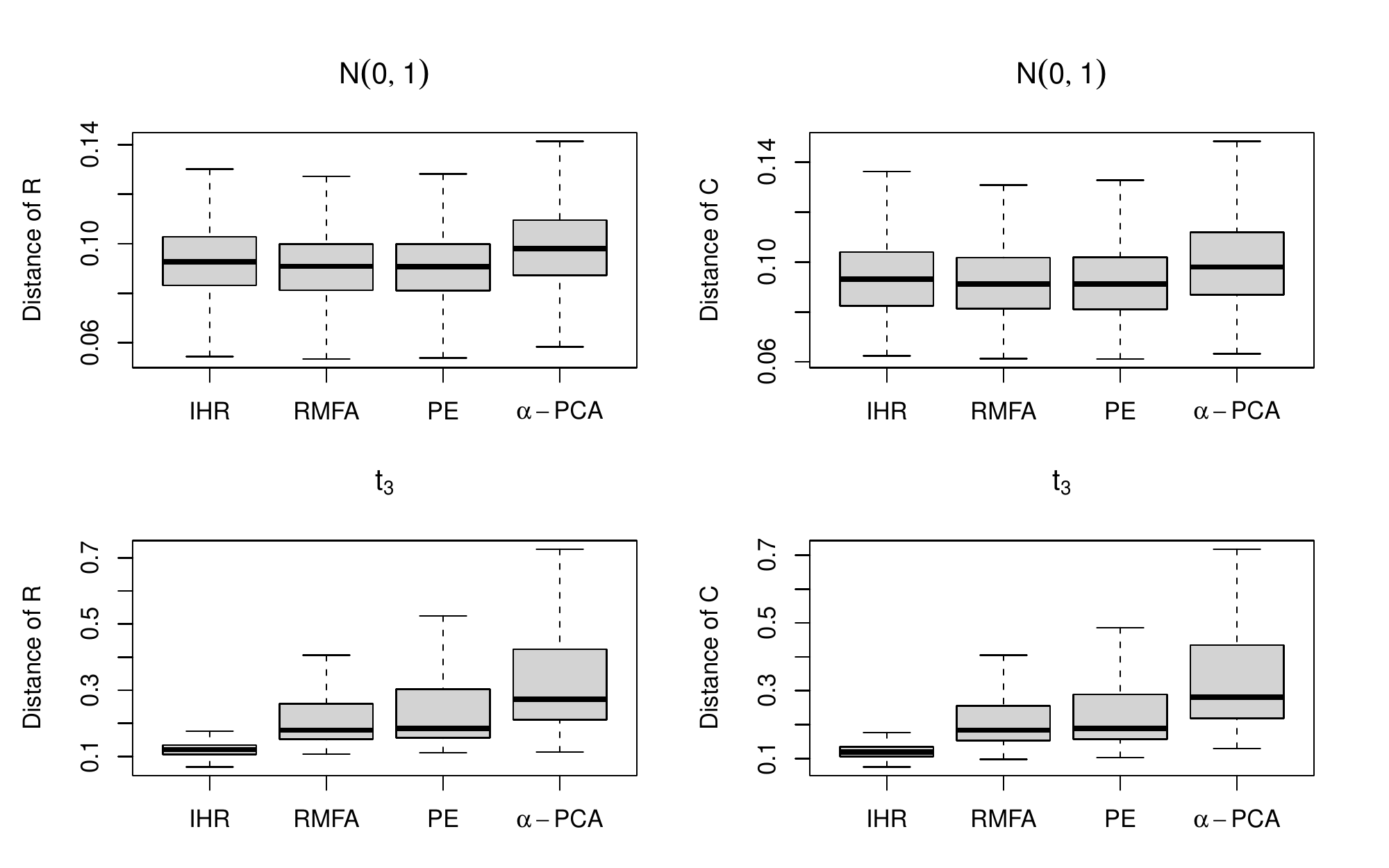}}
    	\caption{\label{fig:1}Boxplots of the distance between the estimated loading spaces and the true loading spaces by IHR, RMFA \citep{He2021MatrixFA}, PE \citep{Yu2021Projected} and $\alpha$-PCA ($\alpha=0$) \citep{fan2021} under $T=p_{1}=p_{2}=20$. The top panels depict the distance $\cD(\hat{\Rb},\Rb_{0})$ and $\cD(\hat{\Cb},\Cb_{0})$ under the element-wise normal distribution of idiosyncratic errors and the bottom panels correspond to the element-wise $t_{3}$ distribution. }
    \end{figure}

In many research areas such as finance and economics, heavy-tailed
data sets are common and it is urgent to develop robust procedures for matrix factor models.  In this paper we consider the element-wise type Huber loss instead of the matrix Frobenius norm type Huber loss by \cite{He2021MatrixFA} and propose an Iterative Huber Regression (IHR) algorithm. We derive the convergence rates of the robust estimators for loadings, factors and common components  under finite second moment assumption of the idiosyncratic errors, which are faster than the rates derived in \cite{He2021MatrixFA}. In addition, we also derive the asymptotic distributions of the estimators under some mild conditions. Two methods based on rank minimization and eigenvalue-ratio are given to estimate the pair of factor numbers consistently.
To check the sensitivity of the PE, $\alpha$-PCA, RMFA and IHR methods to the tail properties of the idiosyncratic errors, we generate the entries of idiosyncratic errors from the standard normal distribution or symmetric $t_{3}$ distribution, see Section \ref{sec:4.1} for detailed data generating mechanism. Figure \ref{fig:1} depicts the boxplots of the row (left panel) and column (right panel) factor loading estimation errors based on 500 replications, from which we can see that all the methods preform almost the same under normal distribution, while the IHR method results in much smaller estimation errors as the distribution tails become heavier.

 The most related literature is  \cite{He2021MatrixFA}, in which the convergence rates of the estimated factor loadings and scores in the sense of the average Frobenius norm is $1/\min\{p_{1},p_{2}\}$. This is much slower than the rate $1/\min\{Tp_{1},Tp_{2},p_{1}p_{2}\}$ derived in this paper. We also derive the limiting distribution of the estimates, which is not discussed in \cite{He2021MatrixFA}. In summary, the contribution of the current work lies in the following aspects: firstly, our work serves as a
much-needed addition to the scarce literature on robust factor analysis for large-dimensional matrix time series; secondly, the proposed algorithm is computationally efficient and the estimates achieve
the convergence rates $1/\min\{Tp_{1},Tp_{2},p_{1}p_{2}\}$ under the finite second-moment condition on the
idiosyncratic errors, which is much faster than that of \cite{He2021MatrixFA}; thirdly,
we also derived the asymptotic distributions of the estimates under mild
conditions, which is the first time for robust matrix factor analysis; fourthly, we introduce both  rank minimization and eigenvalue-ratio estimates to  determining the number of factors, which are shown to be consistent and
also complements the scarce literature on robust determination of the row/column factor numbers and is of independent
interest. Finally, we have developed an R package, ``HDMFA", which implements related robust matrix factor
analysis methods found in the literature and is available on CRAN \footnote{\url{https://cran.r-project.org/web/packages/HDMFA/index.html}}.

    The rest of the paper is organized as follows. In section 2, we introduce the model setup and give the iterative Huber regression algorithm  to estimate the loadings and factor matrices. Section 3 presents some technical assumptions and establishes the convergence rates of the theoretical minimizer, the consistency of model  selection criterions and the limiting distributions of the estimators of loadings. Section 4 conducts thorough simulation studies to investigate the finite sample performances of the proposed methods and Section 5 verifies the practical usefulness of our proposed methods by real  data analysis in financial and macroeconomic areas. Detailed proofs of the main theorems are included in the Appendix.

	To end this section, we introduce some of the notations that will be adopted thorughout the article.
	For a matrix $\Ab$, let $\text{tr}(\Ab)$ denote the trace of $\Ab$ and $\lambda_{j}$ denote the $j$th largest eigenvalue of a nonnegative definitive matrix $\Ab$, let $\|\Ab\|_{F}$ denote the Frobenius norm of $\Ab$ and $\|\Ab\|_{2}$ be the spectral norm of matrix $\Ab$, $\|\Ab\|_{\max}$ be the maximum of $|A_{ij}|$. Let $I(\cdot)$ be the indicator function. For a real number $a$, denote $\lfloor a \rfloor$ as the maximum integer smaller than or equal to $a$. Let $\text{sgn}(a)=1$ if $a \geq 0$ and $\text{sgn}(a)=-1$ if $a<0$. Let $A_{jj}$ be the $j$th diagonal element of a square matrix $\Ab$. We define $\text{sgn}(\Ab)$ as a diagonal matrix whose $j$th diagonal element is equal to $\text{sgn}(A_{jj})$. Let $[T]$ denote the set $\{1,\dots\,T\}$. The notation $\xrightarrow{d}$ represents convergence in distribution and $\otimes$ denotes the Kronecker product. For two random series $X_{n}$ and $Y_{n}$, $X_{n} \lesssim Y_{n}$ means that $X_{n}=O_{p}(Y_{n})$ and $X_{n} \gtrsim Y_{n}$ means that $Y_{n}=O_{p}(X_{n})$. The notation $X_{n} \asymp Y_{n}$ means that $X_{n} \lesssim Y_{n}$ and $X_{n} \gtrsim Y_{n}$. The constant $c$ or $C$ may not be identical in different lines.

	\section{Methodology}

In this section, we review the matrix factor model and introduce the iterative Huber regression algorithm  to estimate the loadings and factor matrices. Let $\{\Xb_{t},1\leq t\leq T\}$ be a sequence of $p_1\times p_2$ random matrices. The corresponding matrix
factor model is given by
\begin{equation}\label{MFM}
		 \Xb_{t}=\Rb_{0}\Fb_{0t}\Cb_{0}^{\top}+\Eb_{t}, \ t=1,\dots,T,
	\end{equation}
  where $\Rb_{0}$ is the $p_{1} \times k_{1}$ row factor loading matrix exploiting the variations of $\Xb_{t}$ across the rows, $\Cb_{0}$ is the $p_{2} \times k_{2}$ column factor loading matrix reflecting the differences in the columns of $\Xb_{t}$, $\Fb_{0t}$ is the $k_{1} \times k_{2}$ common factor matrix and $\Eb_{t}$ is the idiosyncratic component. In this section, we first assume that the pair of the factor numbers $k_1,k_2$ are known as a priori and we will discuss how to determine them in Section \ref{sec:facnum}.
The element-wise Huber loss formulation is given as follows:
    \begin{equation}\label{equ:opt}
    \begin{aligned}
    \min \limits_{\Rb,\Cb,\Fb_{t}} L^{H}(\Rb,\Cb,\Fb_{t})&=\min \limits_{\Rb,\Cb, \Fb_{t}} \dfrac{1}{Tp_{1}p_{2}}\sum_{t=1}^{T}\sum_{i=1}^{p_{1}}\sum_{j=1}^{p_{2}}H_{\tau}\left(x_{t,ij}-\br_{i}^{\top}\Fb_{t}\bc_{j}\right),\\
    & \text{s.t.}	\ \ \dfrac{1}{p_{1}}\Rb^{\top}\Rb=\Ib_{k_{1}}, 	 \dfrac{1}{p_{2}}\Cb^{\top}\Cb=\Ib_{k_{2}},
    \end{aligned}
    \end{equation}
    where the constraints are for the identification of the factor model and correspond to an assumption that the common factors are
``strong" or ``pervasive" across both the row and column dimensions, see also \cite{He2021MatrixFA}.
 In the remainder of this section,  we introduce an Iterative Huber Regression (IHR) algorithm to solve the optimization problem in (\ref{equ:opt}).

Let $\Rb^{\top}=(\br_{1}, \dots, \br_{p_{1}})$, $\Cb^{\top}=(\bc_{1}, \dots, \bc_{p_{2}})$, $\btheta=(\br_{1}^{\top}, \dots, \br_{p_{1}}^{\top}; \bc_{1}^{\top}, \dots, \bc_{p_{2}}^{\top}; \text{Vec}(\Fb_{1})^{\top}, \dots, \text{Vec}(\Fb_{T})^{\top})^{\top}$ and $\btheta_{0}=(\br_{01}^{\top}, \dots, \br_{0p_{1}}^{\top}; \bc_{01}^{\top}, \dots, \bc_{0p_{2}}^{\top}; \text{Vec}(\Fb_{01})^{\top}, \dots, \text{Vec}(\Fb_{0T})^{\top})^{\top}$ be the true parameters, and $N=p_{1}k_{1}+p_{2}k_{2}+Tk_{1}k_{2}$. For ease of presentation, we denote the objective function in (\ref{equ:opt}) as
$$\MM_{Tp_{1}p_{2}}(\btheta)=\dfrac{1}{Tp_{1}p_{2}}\sum_{t=1}^{T}\sum_{i=1}^{p_{1}}\sum_{j=1}^{p_{2}}H_{\tau}\left(x_{t,ij}-\br_{i}^{\top}\Fb_{t}\bc_{j}\right).$$
and the theoretical minimizer  as $\hat{\btheta}=(\hat{\br}_{1}^{\top}, \dots, \hat{\br}_{p_{1}}^{\top}; \hat{\bc}_{1}^{\top}, \dots, \hat{\bc}_{p_{2}}^{\top}; \text{Vec}(\hat{\Fb}_{1})^{\top}, \dots, \text{Vec}(\hat{\Fb}_{T})^{\top})^{\top}$, i.e., $$\hat{\btheta}=\argmin \limits_{\btheta \in \bTheta}\MM_{Tp_{1}p_{2}}(\btheta),$$
where
$$
\begin{array}{lll}
\bTheta=&\bigg\{\btheta\in \RR^{N}: \br_{i} \in \cR \subset \RR^{k_{1}}, \bc_{j} \in \cC \subset \RR^{k_{2}}, \Fb_{t} \in \cF \subset \RR^{k_{1}\times k_{2}}  \ \ \text{for all} \  \ i, j, t,\\
 &\dfrac{1}{p_{1}}\Rb^{\top}\Rb=\Ib_{k_{1}}, \dfrac{1}{p_{2}}\Cb^{\top}\Cb=\Ib_{k_{2}}, \dfrac{1}{T}\sum_{t=1}^{T}\Fb_{t}\Fb_{t}^{\top}=\bSigma_{1}, \dfrac{1}{T}\sum_{t=1}^{T}\Fb_{t}^{\top}\Fb_{t}=\bSigma_{2} \bigg\},
 \end{array}$$
and $\bSigma_{i}$ is a $k_{i} \times k_{i}$ diagonal matrix, $i=1,2$.
We further define the following element-wise objective functions for ease of clarification,
$$\MM_{i,Tp_{2}}(\br,\Fb_{t},\Cb)=\dfrac{1}{Tp_{2}}\sum_{t=1}^{T}\sum_{j=1}^{p_{2}}H_{\tau}(x_{t,ij}-\br^{\top}\Fb_{t}\bc_{j}),$$
$$\MM_{j,Tp_{1}}(\Rb,\Fb_{t},\bc)=\dfrac{1}{Tp_{1}}\sum_{t=1}^{T}\sum_{i=1}^{p_{1}}H_{\tau}(x_{t,ij}-\br_{i}^{\top}\Fb_{t}\bc),$$
$$\MM_{t,p_{1}p_{2}}(\Rb,\text{Vec}(\Fb),\Cb)=\dfrac{1}{p_{1}p_{2}}\sum_{i=1}^{p_{1}}\sum_{j=1}^{p_{2}}H_{\tau}(x_{t,ij}-(\bc_{j}\otimes\br_{i})^{\top}\text{Vec}(\Fb)),$$
and clearly we have that $$\MM_{Tp_{1}p_{2}}(\btheta)=p_{1}^{-1}\sum_{i=1}^{p_{1}}\MM_{i,Tp_{2}}(\br,\Fb_{t},\Cb)=p_{2}^{-1}\sum_{j=1}^{p_{2}}\MM_{j,Tp_{1}}(\Rb,\Fb_{t},\bc)=T^{-1}\sum_{t=1}^{T}\sum_{t=1}^{T}\MM_{t,p_{1}p_{2}}(\Rb,\text{Vec}(\Fb),\Cb).$$ Although the objective function $\MM_{Tp_{1}p_{2}}(\btheta)$ is in general non-convex jointly
in all parameters, $\MM_{i,Tp_{2}}(\br,\Fb_{t},\Cb)$ is convex in $\br$ for every $i$  given $\{\Fb_{t}\}_{t=1}^{T}$ and $\Cb$,  $\MM_{j,Tp_{1}}(\Rb,\Fb_{t},\bc)$ is convex in $\bc$ for each $j$ given $\{\Fb_{t}\}_{t=1}^{T}$ and $\Rb$ and $\MM_{t,p_{1}p_{2}}(\Rb,\text{Vec}(\Fb),\Cb)$ is also convex in $\text{Vec}(\Fb)$   for each $t$ given $\Rb$ and $\Cb$. The above
fact motivates one to minimize the objective function $\MM_{Tp_{1}p_{2}}(\btheta)$ alternatively over $\Rb,\Cb,\text{Vec}(\Fb)$, each time optimizing one
argument while keeping the other two fixed.

To ensure that $\Rb, \Cb, \Fb_{t}$ satisfy the identification condition, one could normalize the loading and factor matrices as follows. Denote  $\tilde{\Rb}^{(s)}, \tilde{\Cb}^{(s)}$ and $\tilde{\Fb}_{t}^{(s)}$ as the estimates of $\Rb_{0}, \Cb_{0}$, and $\Fb_{0t}$ in the $s$-th step, respectively. Then perform singular value decomposition to matrices $\tilde{\Rb}^{(s)}$ and $\tilde{\Cb}^{(s)}$ and obtain
$$\tilde{\Rb}^{(s)}=\Ub_{\tilde{\Rb}^{(s)}}\bLambda_{\tilde{\Rb}^{(s)}}\Vb_{\tilde{\Rb}^{(s)}}^{\top}=\Ub_{\tilde{\Rb}^{(s)}}\Qb_{\tilde{\Rb}^{(s)}}, \ \  \tilde{\Cb}^{(s)}=\Ub_{\tilde{\Cb}^{(s)}}\bLambda_{\tilde{\Cb}^{(s)}}\Vb_{\tilde{\Cb}^{(s)}}^{\top}=\Ub_{\tilde{\Cb}^{(s)}}\Qb_{\tilde{\Cb}^{(s)}},$$
where $\Qb_{\tilde{\Rb}^{(s)}}=\bLambda_{\tilde{\Rb}^{(s)}}\Vb_{\tilde{\Rb}^{(s)}}^{\top}$ and $\Qb_{\tilde{\Cb}^{(s)}}=\bLambda_{\tilde{\Cb}^{(s)}}\Vb_{\tilde{\Cb}^{(s)}}^{\top}$.
Further define two covariance matrices as $$\tilde{\bSigma}_{1}^{(s)}=\dfrac{1}{Tp_{1}p_{2}}\sum_{t=1}^{T}\Qb_{\tilde{\Rb}^{(s)}}\tilde{\Fb}_{t}^{(s)}\tilde{\Cb}^{(s)\top}\tilde{\Cb}^{(s)}\tilde{\Fb}_{t}^{(s)\top}\Qb_{\tilde{\Rb}^{(s)}}^{\top}, \ \ \tilde{\bSigma}_{2}^{(s)}=\dfrac{1}{Tp_{1}p_{2}}\sum_{t=1}^{T}\Qb_{\tilde{\Cb}^{(s)}}\tilde{\Fb}_{t}^{(s)\top}\tilde{\Rb}^{(s)\top}\tilde{\Rb}^{(s)}\tilde{\Fb}_{t}^{(s)}\Qb_{\tilde{\Cb}^{(s)}}^{\top},$$
and denote their eigenvalue decomposition as
$$\tilde{\bSigma}_{1}^{(s)}=\tilde{\bGamma}_{1}^{(s)}\tilde{\bLambda}_{1}^{(s)}\tilde{\bGamma}_{1}^{(s)\top}, \ \ \tilde{\bSigma}_{2}^{(s)}=\tilde{\bGamma}_{2}^{(s)}\tilde{\bLambda}_{2}^{(s)}\tilde{\bGamma}_{2}^{(s)\top}.$$
Finally we can obtain the normalized loading and factor matrices as
\begin{equation}\label{Normalized R}
\hat{\Rb}^{(s)}=\sqrt{p_{1}}\Ub_{\tilde{\Rb}^{(s)}}\tilde{\bGamma}_{1}^{(s)},
\end{equation}
\begin{equation}\label{Normalized C}
\hat{\Cb}^{(s)}=\sqrt{p_{2}}\Ub_{\tilde{\Cb}^{(s)}}\tilde{\bGamma}_{2}^{(s)},
\end{equation}
\begin{equation}\label{Normalized F}
\hat{\Fb}_{t}^{(s)}=\dfrac{1}{\sqrt{p_{1}p_{2}}}\tilde{\bGamma}_{1}^{(s)\top}\Qb_{\tilde{\Rb}^{(s)}}\tilde{\Fb}_{t}^{(s)}\Qb_{\tilde{\Cb}^{(s)}}^{\top}\tilde{\bGamma}_{2}^{(s)}.
\end{equation}

We iteratively estimate the loading and factor matrices by the Huber regression. Once we get an estimate in the iterative procedure, we perform the normalization procedure described as above. The normalization step would not change the estimate of the common components. We name the above algorithm as  Iterative Huber Regression  (IHR)  algorithm, which is computationally efficient. In each iteration, the Huber regression can be implemented by R package  ``MASS" and  the iteration is terminated until the common components of two adjacent steps satisfy $\sum_{t=1}^{T}\|\hat{\Sbb}_{t}^{(s+1)}-\hat{\Sbb}_{t}^{(s)}\|_{F} \leq 10^{-4} \cdot Tp_{1}p_{2}$. In the following, we briefly discuss the determination of tuning parameter $\tau$ in the Huber loss. In essence, the Huber regression can be reformulated as a weighted least squares problem.
For better illustration, we consider the general optimization problem for robust regression, that is,  $\min_{\bbeta} \sum_{i=1}^n H_{\tau}(y_i-\bz_i^{\top}\bbeta)$. For any index $i$ such that $y_i-\bz_i^{\top}\bbeta \neq 0$, the necessary condition of the optimization is that
\begin{equation}\label{equ:hnc}
	 \sum_{i=1}^{n}H^{(1)}_{\tau}(y_i-\bz_i^{\top}\bbeta)\bz_i=\sum_{i=1}^n \frac{H^{(1)}_{\tau}(y_i-\bz_i^{\top}\bbeta)}{(y_i-\bz_i^{\top}\bbeta)}\bz_i(y_i-\bz_i^{\top}\bbeta):= \sum_{i=1}^n \omega_i\bz_i(y_i-\bz_i^{\top}\bbeta)=0,
\end{equation}
which can also be viewed as the KKT condition of weighted least squares. If $|y_i-\bz_i^{\top}\bbeta| \leq \tau$, then $\omega_i=1$, hence, to determine the tuning parameter $\tau$ is equivalent to choose weights $\omega_i$. In simulations, the $\hat{\omega}^{(s)}$ is updated by the re-weighted least square and then obtain the $\hat{\bbeta}^{(s)}$ in the $s$-th iteration. Specifically, we set $$\omega^{(s+1)}_i=\min\rbr{1,\frac{c_1}{|\hat{\epsilon}^{(s)}_i| \cdot c_2}},$$ where $\widehat{\epsilon}^{(s)}_i=y_i-\bz_i^{\top}\widehat{\bbeta}^{(s)}$ for the given $\hat{\beta}^{(s)}$, $c_1$ and $c_2$ are scalar parameters. We choose $\hat{\beta}^{(0)}$ as the least square estimator, and set the scalar parameters $c_1=1.345$ and $c_2=0.6745/\text{Median}(|\hat{\epsilon}_i^{(s)}|)$, where $\text{Median}(|\hat{\epsilon}_i^{(s)}|)$ is the sample median of $\{|\hat{\epsilon}_i^{(s)}|\}$. Finally,
 the detailed procedures of IHR are summarized in Algorithm \ref{alg:3}.

 \begin{algorithm}
	 \renewcommand{\algorithmicrequire}{\textbf{Input:}}
	 \renewcommand{\algorithmicensure}{\textbf{Output:}}
	\caption{Iterative Huber Regression (IHR) algorithm for matrix factor model}
	\label{alg:3}
	\begin{algorithmic}[1]
	\Require Data matrices $\{\Xb_{t}\},t \leq T$, the pair of estimated row and column factor numbers $k_{1}$ and $k_{2}$, the initial estimators $\hat{\Rb}^{(0)},\hat{\Cb}^{(0)}$
	\Ensure Factor loading matrices $\hat{\Rb}, \hat{\Cb}$ and factor matrix $\hat{\Fb}_{t},t \leq T$
	\State  normalize $\hat{\Rb}^{(0)}, \hat{\Cb}^{(0)}$ such that they satisfy the identification condition and still denote as $\hat{\Rb}^{(0)}, \hat{\Cb}^{(0)}$; obtain the initial estimator $\hat{\Fb}_{t}^{(1)}$ by  $\hat{\Fb}_{t}^{(1)}=1/(p_{1}p_{2})\hat{\Rb}^{(0)\top}\Xb_{t}\hat{\Cb}^{(0)}$;
	\State given $\{\hat{\Fb}_{t}^{(1)}\}_{t=1}^{T}$ and $\hat{\Cb}^{(0)}$, solve $\tilde{\br}_{i}^{(1)}=\argmin_{\br}\MM_{i,Tp_{2}}(\br,\hat{\Fb}_{t}^{(1)},\hat{\Cb}^{(0)})$ for $i=1,\dots,p_{1}$; given $\{\hat{\Fb}_{t}^{(1)}\}_{t=1}^{T}$ and $\tilde{\Rb}^{(1)}$, solve $\tilde{\bc}_{j}^{(1)}=\argmin_{\bc}\MM_{j,Tp_{1}}(\tilde{\Rb}^{(1)},\hat{\Fb}_{t}^{(1)},\bc)$ for $j=1,\dots,p_{2}$; given $\tilde{\Rb}^{(1)}, \tilde{\Cb}^{(1)}$, solve  $\text{Vec}(\tilde{\Fb}_{t}^{(2)})=\argmin_{\text{Vec}(\Fb)}\MM_{t,p_{1}p_{2}}(\tilde{\Rb}^{(1)},\text{Vec}(\Fb),\tilde{\Cb}^{(1)})$ for $t=1,\dots,T$;
	\State normalize $\tilde{\Rb}^{(1)}, \tilde{\Cb}^{(1)}$ and $\tilde{\Fb}_{t}^{(2)}$ so that they satisfy the identifiability conditions by Equation (\ref{Normalized R}), by Equation (\ref{Normalized C})and by Equation (\ref{Normalized F}), respectively; and denote as $\hat{\Rb}^{(1)},\hat{\Cb}^{(1)}, \hat{\Fb}_{t}^{(2)}$;
	\State repeat step 2-3 until convergence and output the estimators from the last step and denoted as $\hat{\Rb}$, $\hat{\Cb}$ and $\hat{\Fb}_{t},t \leq T$, respectively.
		
	\end{algorithmic}
\end{algorithm}

\section{Theoretical Properties}
In this section, we establish the convergence rates for the estimators of factor loadings,  factor scores and also the common components. We propose two methods to estimate the row and column factor numbers, one is based on rank-minimization and the other is based on eigenvalue-ratio. At last, we give the asymptotic normality of the estimators for factor loadings. To establish the theoretical results, we first introduce some assumptions which are mild/common in the related literature.

\begin{asmp}\label{Assumption 1}
$\cR, \cC, \cF$ are compact sets and $\btheta_{0} \in \bTheta$. The factor matrices satisfy $$\dfrac{1}{T}\sum_{t=1}^{T}\Fb_{0t}\Fb_{0t}^{\top} = \text{diag}(\sigma_{1,1}^{T}, \dots, \sigma_{1,k_{1}}^{T}) \ \text{and} \ \dfrac{1}{T}\sum_{t=1}^{T}\Fb_{0t}^{\top}\Fb_{0t} = \text{diag}(\sigma_{2,1}^{T}, \dots, \sigma_{2,k_{2}}^{T}),$$
where $\sigma_{i,1}^{T} \geq \dots \geq \sigma_{i,k_{i}}^{T}$ and $\sigma_{i,j}^{T} \to \sigma_{i,j}$ as $T \to \infty$ for $j=1, \dots, k_{i}$ with $\infty > \sigma_{i,1} > \cdots > \sigma_{i,k_{i}} > 0,i=1,2.$

\end{asmp}

\begin{asmp}\label{Assumption 2}
Given $\{\Fb_{0t}, 1 \leq t \leq T\}$, $\{e_{t,ij}, 1 \leq t \leq T, 1 \leq i \leq p_{1}, 1 \leq j \leq p_{2}\}$ are independent across $t, i$ and $j$.
\end{asmp}

\begin{asmp}\label{Assumption 3}
	
(1) The distribution functions of $e_{t,ij}$ given $\{\Fb_{0t}\}$ have common support covering an open neighborhood of the origin and its conditional density function, denoted as $f_{t,ij}$, is continuous and symmetric with respect to the origin;
(2) Suppose for any compact set $\cE \subset \RR$ and any $e \in \cE$, there exists a positive constant $\underline{f}>0$ (depending on $\cE$) such that $f_{t,ij}(e) \geq \underline{f}$ for all $t, i, j$.
(3) $\EE\left(e_{t,ij}^{2}|\{\Fb_{0t}\}\right) \leq C$ for some constant $C>0$.
\end{asmp}

 Assumption \ref{Assumption 1} is a standard strong/pervasive factor condition and we refer, for example, to \cite{fan2021} and \cite{He2021MatrixFA}.  Assumption \ref{Assumption 2} requires that given the factor matrices $\{\Fb_{0t}\}_{t=1}^{T}$, the idiosyncratic errors are independent across the  row, column and time series dimensions, see also \cite{He2021MatrixFA}.
 Assumption \ref{Assumption 3} exerts some conditions on the conditional distribution
of the idiosyncratic errors given $\{\Fb_{0t}\}_{t=1}^{T}$, and  Assumption \ref{Assumption 3} (1)-(2) are similar to the Assumption 1 (ii) and 1 (iii) in \cite{Chen2021Quantile} and Assumption C1 and C2 in \cite{ando2020quantile}. In Assumption \ref{Assumption 3} (2), we also assume the conditional density function is symmetric with respect to the origin, thus, we have $\EE(H_{\tau}^{(1)}(e_{t,ij})|\{\Fb_{0t}\})=0$, where $H_{\tau}^{(1)}(e_{t,ij})$ is the derivative function of $H_{\tau}(\cdot)$ evaluated at $e_{t,ij}$. Assumption \ref{Assumption 3} (3) imposes a finite second moment of $e_{t,ij}$ conditional on $\{\Fb_{0t}\}_{t=1}^{T}$, which relaxes the typical finite fourth or even eighth moment assumption in the literature \citep{bai2003inferential,fan2021,Yu2021Projected}.

\subsection{Convergence rates}
In this section, we present the theoretical results on convergence rates of the proposed estimators. To this end,
first denote $\hat{\Rb}^{\top}=(\hat{\br}_{1}, \dots, \hat{\br}_{p_{1}})$ and $\hat{\Cb}^{\top}=(\hat{\bc}_{1}, \dots, \hat{\bc}_{p_{2}})$, $\Rb_{0}^{\top}=(\br_{01}, \dots, \br_{0p_{1}}), \Cb_{0}=(\bc_{01}, \dots, \bc_{0p_{2}})$. The following theorem establishes the average convergence rate of $\hat{\Rb}, \hat{\Cb}$ and $\hat{\Fb}_{t}$ in terms of matrix Fronbenius norm.
\begin{theorem}\label{Theorem1}
Under Assumptions \ref{Assumption 1}-\ref{Assumption 3}, let $\hat{\Hb}_{1}=\text{sgn}(\Rb_{0}^{\top}\hat{\Rb}/p_{1})$, $\hat{\Hb}_{2}=\text{sgn}(\Cb_{0}^{\top}\hat{\Cb}/p_{2})$ and $\tau$ be a fixed positive constant, then we have $$\dfrac{1}{p_{1}}\|\hat{\Rb}-\Rb_{0}\hat{\Hb}_{1}\|_{F}^{2}+\dfrac{1}{p_{2}}\|\hat{\Cb}-\Cb_{0}\hat{\Hb}_{2}\|_{F}^{2}+\dfrac{1}{T}\sum_{t=1}^{T}\|\hat{\Fb}_{t}-\hat{\Hb}_{1}\Fb_{0t}\hat{\Hb}_{2}\|_{F}^{2}=O_{p}\left(\dfrac{1}{D^{2}}\right),$$
where $D=\min\{\sqrt{Tp_{1}},\sqrt{Tp_{2}},\sqrt{p_{1}p_{2}}\}.$
\end{theorem}

Theorem \ref{Theorem1} shows that $\|\hat{\Rb}-\Rb_{0}\hat{\Hb}_{1}\|_{F}/\sqrt{p_{1}}=O_{p}\left(D^{-1}\right)$, $\|\hat{\Cb}-\Cb_{0}\hat{\Hb}_{2}\|_{F}/\sqrt{p_{2}}=O_{p}\left(D^{-1}\right)$ and $\|\hat{\Fb}_{t}-\hat{\Hb}_{1}\Fb_{0t}\hat{\Hb}_{2}\|_{F}/\sqrt{T}=O_{p}\left(D^{-1}\right)$ for any $t$. The convergence rate is the same as that derived in Theorem 3.1 of \cite{He2021MatrixFA}, while \cite{He2021MatrixFA} relies on the  sub-Gaussianity of the idiosyncratic errors. Under the same finite second moments assumption, the rate derived here is much faster than the rate $\min\{p_{1},p_{2}\}$ in Theorem 4.1 of \cite{He2021MatrixFA}.

\subsection{Determining the pair of factor numbers $k_{1}$ and $k_{2}$}\label{sec:facnum}
The pair of factor numbers $k_{1}, k_{2}$ need to be determined before we estimate the factor loadings and scores.

Assume  $m_{1} > k_{1}, m_{2} > k_{2}$, and let $\Rb^{m_{1}\top}=(\br_{1}^{m_{1}}, \dots, \br_{p_{1}}^{m_{1}}), \Cb^{m_{2}\top}=(\bc_{1}^{m_{2}}, \dots, \bc_{p_{2}}^{m_{2}}).$
Denote $\btheta^{m}=(\br_{1}^{m_{1}\top}, \dots, \br_{p_{1}}^{m_{1}\top}; \bc_{1}^{m_{2}\top}, \dots, \bc_{p_{2}}^{m_{2}\top}; \text{Vec}(\Fb_{1}^{m\top}), \dots, \text{Vec}(\Fb_{T}^{m\top}))^{\top}$, where $\br_{i}^{m_{1}} \in \RR^{m_{1}}, \bc_{j}^{m_{2}} \in \RR^{m_{2}}, \Fb_{t}^{m} \in \RR^{m_{1} \times m_{2}}$, for all $i, j, t$.

Consider the following identifiability condition:
\begin{equation}\label{equ:identi}
\begin{array}{ccc}
\dfrac{1}{p_{1}}\Rb^{m_{1}\top}\Rb^{m_{1}}=\Ib_{m_{1}}, \ \  \dfrac{1}{p_{2}}\Cb^{m_{2}\top}\Cb^{m_{2}}=\Ib_{m_{2}}, \ \
\dfrac{1}{T}\sum_{t=1}^{T}\Fb_{t}^{m}\Fb_{t}^{m\top} \ \text{and} \ \dfrac{1}{T}\sum_{t=1}^{T}\Fb_{t}^{m\top}\Fb_{t}^{m} \\ \ \text{are diagonal matrices with non-increasing diagonal elements}.
\end{array}
\end{equation}

Let $\cR^{m_{1}}, \cC^{m_{2}}, \cF^{m}$ be compact subsets of $\RR^{m_{1}}, \RR^{m_{2}}, \RR^{m_{1} \times m_{2}}$, respectively. Assume that $(\br_{0i}^{\top},\bm{0}_{1 \times (m_{1}-k_{1})})^{\top} \in \cR^{m_{1}}, (\bc_{j0}^{\top}, \bm{0}_{1 \times (m_{2}-k_{2})}) \in \cC^{m_{2}}$, and $$\begin{pmatrix}
\Fb_{0t} & \b0_{k_{1} \times (m_{2}-k_{2})} \\ \bm{0}_{(m_{1}-k_{1}) \times k_{2}} & \b0_{(m_{1}-k_{1}) \times (m_{2}-k_{2})}
\end{pmatrix} \in \cF^{m}, \ \text{for all}\  i, j, t.$$
Define $$\hat{\btheta}^{m}=\argmin_{\btheta^{m} \in \bTheta^{m}}=\dfrac{1}{Tp_{1}p_{2}}\sum_{t=1}^{T}\sum_{i=1}^{p_{1}}\sum_{j=1}^{p_{2}}H_{\tau}(x_{t,ij}-\br_{i}^{m_{1}\top}\Fb_{t}^{m}\bc_{j}^{m_{2}}),$$
where $\bTheta^{m}:=\{\btheta^{m}:\br_{i}^{m_{1}} \in \cR^{m_{1}}, \bc_{j}^{m_{2}} \in \cC^{m_{2}}, \Fb_{t}^{m} \in \cF^{m}, \ \text{and} \ \Rb^{m_{1}}, \Cb^{m_{2}}, \Fb_{t}^{m} \ \text{satisfy (\ref{equ:identi})}\}.$ Moreover, denote $$\hat{\bSigma}_{1}=\dfrac{1}{T}\sum_{t=1}^{T}\hat{\Fb}_{t}^{m}\hat{\Fb}_{t}^{m\top}=\text{diag}(\hat{\sigma}_{1,1},\dots,\hat{\sigma}_{1,m_{1}}), \ \hat{\bSigma}_{2}=\dfrac{1}{T}\sum_{t=1}^{T}\hat{\Fb}_{t}^{m\top}\hat{\Fb}_{t}^{m}=\text{diag}(\hat{\sigma}_{2,1},\dots,\hat{\sigma}_{2,m_{2}}).$$
The rank minimization estimators for the pair of factor numbers are
\begin{equation}\label{equ:rankfac}
\hat{k}_{1}^{RM}=\sum_{i=1}^{m_{1}}\bm{1}\{\hat{\sigma}_{1,i} > P_{1}\}, \ \ \hat{k}_{2}^{RM}=\sum_{j=1}^{m_{2}}\bm{1}\{\hat{\sigma}_{2,j} > P_{2}\},
\end{equation}
where $P_{1}$ and $P_{2}$ are sequences that go to 0 as $T, p_{1}, p_{2} \to \infty$.
The following theorem establishes the consistency of the estimators $\hat{k}_{1}^{RM}$ and $\hat{k}_{2}^{RM}$.

\begin{theorem}\label{the number of factors RK}
Under Assumptions \ref{Assumption 1}-\ref{Assumption 3}, if $m_{1} > k_{1} > 0, m_{2} > k_{2} > 0$, $P_{1} \to 0$, $P_{2} \to 0$ and $P_{1}D^{2} \to \infty$, $P_{2}D^{2} \to \infty$, then we have $\PP(\hat{k}_{1}^{RM}=k_{1}) \to 1$ and $\PP(\hat{k}_{2}^{RM}=k_{2}) \to 1$, as $T, p_{1}, p_{2} \to \infty$.
\end{theorem}

For pre-determined $m_{1}$ and $m_{2}$ such that  $m_{1} > k_{1}, m_{2} > k_{2}$, the eigenvalue-ratio based estimators of $k_1$ and $k_2$ are defined as
$$\hat{k}_{1}^{ER}=\argmax \limits_{i \leq m_{1}-1}\dfrac{\lambda_{i}(\hat{\bSigma}_{1})}{\lambda_{i+1}(\hat{\bSigma}_{1})+c\alpha},  \ \ \ \hat{k}_{2}^{ER}=\argmax \limits_{j \leq m_{2}-1}\dfrac{\lambda_{j}(\hat{\bSigma}_{2})}{\lambda_{j+1}(\hat{\bSigma}_{2})+c\alpha},$$
where $\alpha=\max\{(Tp_{1})^{-1}, (Tp_{2})^{-1}, (p_{1}p_{2})^{-1}\}$ is the convergence rate derived in Theorem \ref{Theorem1} and $c\alpha$ is the lower bound of the denominator. We set $c=10^{-4}$ in empirical study.  The advantage of the eigenvalue-ratio  method compared with the rank minimization method lies in that  we do not need to determine the threshold value $P_{i}$. The following theorem establishes the consistency of the estimators $\hat{k}_{1}^{ER}$ and $\hat{k}_{2}^{ER}$.
\begin{theorem}\label{the number of factors ER}
Under Assumptions \ref{Assumption 1}-\ref{Assumption 3}, when $T, p_{1}, p_{2} \to \infty$, $m_{1}$ and $m_{2}$ are predetermined constants with $m_{1} > k_{1} > 0, m_{2} > k_{2} > 0$, then we have $\PP(\hat{k}_{1}^{ER}=k_{1}) \to 1$ and $\PP(\hat{k}_{2}^{ER}=k_{2}) \to 1$.
\end{theorem}

\subsection{Asymptotic distribution of estimators}
In this section, we establish the asymptotic normality of the estimators for factor loadings so that statistical inference is feasible. We introduce some notations first.
Define $$\bPhi_{i}=\lim\limits_{Tp_{2} \to \infty}\dfrac{1}{T,p_{2}}\sum_{t=1}^{T}\sum_{j=1}^{p_{2}}\int_{-\tau}^{\tau}f_{t,ij}(e) \, de \cdot \Fb_{0t}\bc_{0j}\bc_{0j}^{\top}\Fb_{0t}^{\top},$$
$$\bPsi_{j}=\lim\limits_{T,p_{1} \to \infty}\dfrac{1}{T,p_{1}}\sum_{t=1}^{T}\sum_{i=1}^{p_{1}}\int_{-\tau}^{\tau}f_{t,ij}(e) \, de \cdot \Fb_{0t}^{\top}\br_{0i}\br_{0i}^{\top}\Fb_{0t}.$$

We then introduce an assumption,
which is crucial in establishing the asymptotic distribution of $\hat\br_{i} (i \leq p_{1})$ and $\hat\bc_{j} (j \leq p_{2})$.
\begin{asmp}\label{Assumption 4}
(1) $\bPhi_{i} >0, \bPsi_{j} >0$ for all $i \leq p_{1}$, $j \leq p_{2}$.

(2) For any compact set $\cE \in \RR$ and any $e \in \cE$, there exists $0 < \underline{f} < \bar{f} < \infty$, (depending on $\cE$) such that $\underline{f} < f_{t,ij}(e) < \bar{f}$ for all $t, i, j$.

(3) $(i)$ $p_{1} \asymp Tp_{2}$ and $p_{1},Tp_{2} \to \infty$.
$(ii)$ $p_{2} \asymp Tp_{1}$ and $p_{2}, Tp_{1} \to \infty$.
\end{asmp}

 Assumption \ref{Assumption 4} (1) assumes that $\bPhi_{i}$ and $\bPsi_{j}$ are positive definite matrices. Assumption \ref{Assumption 4} (2) assumes the density function is bounded on a compact  domain of definition. Assumption \ref{Assumption 4} (3) gives the scaling requirement on $p_{1}, p_{2}$ and $T$.
The following theorem establishes the asymptotic distributions of $\hat{\br}_{i}$ and $\hat{\bc}_{j}$.
\begin{theorem}\label{Theorem 4}
Suppose Assumptions \ref{Assumption 1}-\ref{Assumption 3} and \ref{Assumption 4} (1)- (2) are satisfied, then the following results hold:

(1) if Assumption \ref{Assumption 4} (3) $(i)$ holds,
$$\sqrt{Tp_{2}}(\hat{\br}_{i}-\hat{\Hb}_{1}\br_{0i}) \xrightarrow{d} N\left(\bm{0}, \bPhi_{i}^{-1}\bSigma_{Tp_{2},i}\bPhi_{i}^{-1}\right) \; \text{for each} \; i \leq p_{1},$$
where $\hat{\Hb}_{1}=\text{sgn}(\Rb_{0}^{\top}\hat{\Rb}/p_{1})$ and
$$\bSigma_{Tp_{2},i}=\lim\limits_{Tp_{2} \to \infty}\dfrac{1}{Tp_{2}}\sum_{t=1}^{T}\sum_{j=1}^{p_{2}}\int_{-\infty}^{\infty}\min\{\tau^{2},e^{2}\}f_{t,ij}(e) \, de \cdot \Fb_{0t}\bc_{0j}\bc_{0j}^{\top}\Fb_{0t}^{\top}.$$

(2) if Assumption \ref{Assumption 4} (3) $(ii)$ holds,
$$\sqrt{Tp_{1}}(\hat{\bc}_{j}-\hat{\Hb}_{2}\bc_{0j}) \xrightarrow{d} N\left(\bm{0},\bPsi_{j}^{-1}\bSigma_{Tp_{1},j}\bPsi_{j}^{-1}\right) \; \text{for each} \; j \leq p_{2},$$
where $\hat{\Hb}_{2}=\text{sgn}(\Cb_{0}^{\top}\hat{\Cb}/p_{2})$ and
$$\bSigma_{Tp_{1},j}=\lim\limits_{Tp_{1} \to \infty}\dfrac{1}{Tp_{1}}\sum_{t=1}^{T}\sum_{i=1}^{p_{1}}\int_{-\infty}^{\infty}\min\{\tau^{2}, e^{2}\}f_{t,ij}(e) \, de \cdot \Fb_{0t}^{\top}\br_{0i}\br_{0i}^{\top}\Fb_{0t}.$$

\end{theorem}

In Theorem \ref{Theorem 4}, the asymptotic variance matrices are unknown and need to be estimated. Intuitively, the
estimators for the asymptotic variance matrices can be naturally constructed as follows:
$$\hat{\bPhi}_{i}=\dfrac{1}{Tp_{2}}\sum_{t=1}^{T}\sum_{j=1}^{p_{2}}H_{\tau}^{(2)}\left(x_{t,ij}-\hat{\br}_{i}^{\top}\hat{\Fb}_{t}\hat{\bc}_{j}\right)\hat{\Fb}_{t}\hat{\bc}_{j} \hat{\bc}_{j}^{\top}\hat{\Fb}_{t}^{\top},$$
$$\hat{\bSigma}_{Tp_{2},i}=\dfrac{1}{Tp_{2}}\sum_{t=1}^{T}\sum_{j=1}^{p_{2}}\left[H_{\tau}^{(1)}\left(x_{t,ij}-\hat{\br}_{i}^{\top}\hat{\Fb}_{t}\hat{\bc}_{j}\right)\right]^{2}\hat{\Fb}_{t}\hat{\bc}_{j} \hat{\bc}_{j}^{\top}\hat{\Fb}_{t}^{\top},$$
and
$$\hat{\bPsi}_{j}=\dfrac{1}{Tp_{1}}\sum_{t=1}^{T}\sum_{i=1}^{p_{1}}H_{\tau}^{(2)}\left(x_{t,ij}-\hat{\br}_{i}^{\top}\hat{\Fb}_{t}\hat{\bc}_{j}\right)\hat{\Fb}_{t}^{\top}\hat{\br}_{i}\hat{\br}_{i}^{\top}\hat{\Fb}_{t},$$
$$\hat{\bSigma}_{Tp_{1},j}=\dfrac{1}{Tp_{1}}\sum_{t=1}^{T}\sum_{i=1}^{p_{1}}\left[H_{\tau}^{(1)}\left(x_{t,ij}-\hat{\br}_{i}^{\top}\hat{\Fb}_{t}\hat{\bc}_{j}\right)\right]^{2}\hat{\Fb}_{t}^{\top}\hat{\br}_{i}\hat{\br}_{i}^{\top}\hat{\Fb}_{t},$$
where $H_{\tau}^{(2)}\left(x\right)=I\{|x| \leq \tau\}$ and  $\left[H_{\tau}^{(1)}\left(x\right)\right]^{2}=x^{2}I\{|x| \leq \tau\}+ \tau^{2} I\{|x| > \tau\}$. Under Assumptions \ref{Assumption 1}-\ref{Assumption 4}, we prove  that the  estimators of the asymptotic covariance matrices are consistent, see the proof in the supplement. Thus we have the following standardized version of Theorem \ref{Theorem 4}.

\begin{theorem}\label{Theorem 5}
Suppose Assumptions \ref{Assumption 1}-\ref{Assumption 3} and \ref{Assumption 4} (1)-(2) are satisfied, then the following results hold:

(1) if Assumption \ref{Assumption 4} (3) $(i)$ holds,
$$\sqrt{Tp_{2}}(\hat{\br}_{i}-\hat{\Hb}_{1}\br_{0i}) \xrightarrow{d} N\left(\bm{0}, \hat\bPhi_{i}^{-1}\hat\bSigma_{Tp_{2},i}\hat\bPhi_{i}^{-1}\right) \; \text{for each} \; i \leq p_{1};$$

(2) if Assumption \ref{Assumption 4} (3) $(ii)$ holds,
$$\sqrt{Tp_{1}}(\hat{\bc}_{j}-\hat{\Hb}_{2}\bc_{0j}) \xrightarrow{d} N\left(\bm{0},\hat\bPsi_{j}^{-1}\hat\bSigma_{Tp_{1},j}\hat\bPsi_{j}^{-1}\right) \; \text{for each} \; j \leq p_{2}.$$
\end{theorem}

At last, we establish the convergence rate of the estimators for common components. To this end, we introduce some notations. Let $\hat{\Sbb}_{t}=\hat{\Rb}\hat{\Fb}_{t}\hat{\Cb}^{\top}$, $\Sbb_{0t}=\Rb_{0}\Fb_{0t}\Cb_{0}^{\top}$ and further define
 $$
 \bDelta_{t}=\lim \limits_{p_{1},p_2 \to \infty}\dfrac{1}{p_{1}p_{2}}\sum_{i=1}^{p_{1}}\sum_{j=1}^{p_{2}}\int_{-\tau}^{\tau}f_{t,ij}(e) \, de \cdot (\bc_{0j} \otimes \br_{0i})(\bc_{0j} \otimes \br_{0i})^{\top}.
 $$
 The following theorem provides the convergence rate of the estimated common components.
\begin{theorem}\label{common components}
Under the Assumptions \ref{Assumption 1}-\ref{Assumption 4} (1), (2), and further assume $\bDelta_{t} > 0$, as $\min\{T,p_{1},p_{2}\} \to \infty$, for any $t \leq T, i \leq p_{1}, j \leq p_{2}$, we have $$|\hat{S}_{t,ij}-S_{0t,ij}|=O_{p}\left(\dfrac{1}{\sqrt{Tp_{1}}}+\dfrac{1}{\sqrt{Tp_{2}}}+\dfrac{1}{\sqrt{p_{1}p_{2}}}\right).$$
\end{theorem}

The convergence rate in Theorem \ref{common components} is the same as that derived in Theorem 3.5 of \cite{Yu2021Projected}, however, we only require the finite second moment of the idiosyncratic errors while \cite{Yu2021Projected} imposes finite eighth moment condition.

\section{Numerical Studies}
In this section, we investigate the empirical performances of the proposed methods by synthetic simulation data.  We first introduce the data generating procedure in Section \ref{sec:4.1}. We  compare the  proposed method with the RMFA method by \citep{He2021MatrixFA}, the PE method by \citep{Yu2021Projected} and the $\alpha$-PCA method by \citep{fan2021} in terms of estimating the loading spaces under different scenarios in Section \ref{sec:4.2}. In Section \ref{sec:4.3}, we compare the finite sample performances of different methods for determining the pair of factor numbers. In Section \ref{sec:4.4}, we numerically verify the asymptotic normality of the estimators for loadings.
	\subsection{Data generating process}\label{sec:4.1}
	The matrix sequences are generated in the following manner.
	We set $k_{1}=3, k_{2}=3$, and draw the entries of $\Rb_{0}$ and $\Cb_{0}$ independently from uniform distribution $U(-1,1)$, and let $\Fb_{0t}$ and $\Eb_{t}$ follow an autoregressive model of order 1, i.e.
	 $$\Fb_{0t}=\phi\Fb_{0(t-1)}+\sqrt{1-\phi^{2}}\bepsilon_{t}, \; \ \  \Eb_{t}=\psi\Eb_{t-1}+\sqrt{1-\psi^{2}}\Ub_{t},$$
	where $\text{Vec}(\bepsilon_{t}) \stackrel{i.i.d}{\sim} \cN(\bm{0}, \Ib_{k_{1} \times k_{2}})$. The entries of $\{\Ub_{t}\}_{t=1}^{T}$ are i.i.d. samples from $\cN(0,1)$ or $t$ distribution with degrees of freedom 3 or 5, including both light-tailed and heavy-tailed scenarios. The parameters $\phi$ and $\psi$ control the temporal correlations. Finally, the observations $\{\Xb_{t}\}_{t=1}^T$ are generated by
	 $\Xb_{t}=\Rb_{0}\Fb_{0t}\Cb_{0}^{\top}+\Eb_{t}$.
	
\begin{table}[htbp]
	\caption{Averaged estimation errors and standard errors of $\cD(\hat{\Rb},\Rb_{0})$ and $ \cD(\hat{\Cb},\Cb_{0})$, for Setting A under $\cN(0,1), t_{5}, t_{3}$ distribution over 500 replications. ``IHR": Iterative Huber Regression. ``RMFM": robust matrix factor analysis method \citep{He2021MatrixFA}. ``PE": projection estimation method \citep{Yu2021Projected}. ``$\alpha$-PCA": $\alpha$-PCA with $\alpha=0$ \citep{fan2021}. } \label{table:1}
	\renewcommand{\arraystretch}{1.2}
	\centering
	\scalebox{0.9}{
		\begin{tabular}{cccccccc}
			\toprule[2pt]
			Evaluation &$T$ &$p_{1}$ &$p_{2}$ &IHR &RMFM &PE &$\alpha$-PCA($\alpha=0$)\\
			\cmidrule(r){1-8}
			\multicolumn{8}{c}{Normal Distribution}\\
			\multirow{5}{2cm}{\centering $\cD(\hat{\Rb},\Rb_{0})$}&
			20 &\multirow{5}{1cm}{\centering 20} &20
			 &0.0938(0.0149)&0.0916(0.0146)&0.0916(0.0146)&0.0992(0.0172)\\
			&50&&50   &0.0363(0.0051)&0.0355(0.0049)&0.0354(0.0049)&0.0388(0.0060)\\
			&100&&100 &0.0181(0.0026)&0.0176(0.0025)&0.0176(0.0025)&0.0193(0.0031)\\
			&150&&150 &0.0118(0.0016)&0.0116(0.0016)&0.0116(0.0016)&0.0126(0.0020)\\
			&200&&200 &0.0090(0.0012)&0.0088(0.0012)&0.0088(0.0012)&0.0096(0.0015)\\
			\cmidrule(r){2-8}
			\multirow{5}{2cm}{\centering $\cD(\hat{\Cb},\Cb_{0})$}&
			20 &\multirow{5}{1cm}{\centering 20} &20
			 &0.0952(0.0171)&0.0929(0.0167)&0.0929(0.0168)&0.1013(0.0201)\\
			&50&&50
			 &0.0578(0.0057)&0.0564(0.0056)&0.0563(0.0056)&0.0581(0.0060)\\
			&100&&100 &0.0405(0.0033)&0.0395(0.0032)&0.0395(0.0032)&0.0401(0.0033)\\
			&150&&150 &0.0328(0.0024)&0.0320(0.0023)&0.0320(0.0023)&0.0323(0.0023)\\
			&200&&200 &0.0287(0.0020)&0.0280(0.0020)&0.0280(0.0020)&0.0282(0.0020)\\
			\hline
			\multicolumn{8}{c}{$t_{5}$ Distribution}\\
			\multirow{5}{2cm}{\centering $\cD(\hat{\Rb},\Rb_{0})$}&
			20 &\multirow{5}{1cm}{\centering 20} &20
			 &0.1086(0.0182)&0.1209(0.0203)&0.1213(0.0205)&0.1426(0.0305)\\
			&50&&50   &0.0413(0.0056)&0.0457(0.0061)&0.0457(0.0061)&0.0544(0.0094)\\
			&100&&100 &0.0203(0.0029)&0.0226(0.0033)&0.0227(0.0033)&0.0271(0.0047)\\
			&150&&150 &0.0134(0.0019)&0.0150(0.0021)&0.0150(0.0021)&0.0179(0.0032)\\
			&200&&200 &0.0102(0.0014)&0.0114(0.0016)&0.0114(0.0016)&0.0137(0.0025)\\
			\cmidrule(r){2-8}
			\multirow{5}{2cm}{\centering $\cD(\hat{\Cb},\Cb_{0})$}&
			20 &\multirow{5}{1cm}{\centering 20} &20
			 &0.1104(0.0196)&0.1227(0.0220)&0.1231(0.0223)&0.1451(0.0344)\\
			&50&&50   &0.0658(0.0066)&0.0728(0.0072)&0.0729(0.0072)&0.0769(0.0084)\\
			&100&&100 &0.0460(0.0038)&0.0510(0.0042)&0.0510(0.0042)&0.0522(0.0045)\\
			&150&&150 &0.0374(0.0027)&0.0414(0.0030)&0.0414(0.0030)&0.0421(0.0031)\\
			&200&&200 &0.0326(0.0023)&0.0361(0.0026)&0.0361(0.0026)&0.0365(0.0026)\\
			\hline
			\multicolumn{8}{c}{$t_{3}$ Distribution}\\
			\multirow{5}{2cm}{\centering $\cD(\hat{\Rb},\Rb_{0})$}&
			20 &\multirow{5}{1cm}{\centering 20} &20
			 &0.1213(0.0206)&0.2463(0.1496)&0.2597(0.1545)&0.3263(0.1460)\\
			&50&&50   &0.0455(0.0066)&0.0982(0.1182)&0.1082(0.1311)&0.1642(0.1274)\\
			&100&&100 &0.0224(0.0033)&0.0417(0.0714)&0.0469(0.0838)&0.0951(0.0969)\\
			&150&&150 &0.0147(0.0020)&0.0323(0.0746)&0.0352(0.0835)&0.0708(0.0863)\\
			&200&&200 &0.0111(0.0015)&0.0207(0.0532)&0.0217(0.0575)&0.0519(0.0657)\\
			\cmidrule(r){2-8}
			\multirow{5}{2cm}{\centering $\cD(\hat{\Cb},\Cb_{0})$}&
			20 &\multirow{5}{1cm}{\centering 20} &20
			 &0.1208(0.0221)&0.2466(0.1491)&0.2589(0.1532)&0.3316(0.1444)\\
			&50&&50   &0.0722(0.0072)&0.1303(0.1121)&0.1399(0.1209)&0.1688(0.1261)\\
			&100&&100 &0.0504(0.0039)&0.0782(0.0665)&0.0852(0.0799)&0.0960(0.0826)\\
			&150&&150 &0.0409(0.0029)&0.0660(0.0706)&0.0711(0.0842)&0.0767(0.0845)\\
			&200&&200 &0.0357(0.0025)&0.0532(0.0511)&0.0555(0.0582)&0.0587(0.0584)\\
			\bottomrule[2pt]
		\end{tabular}
	}
\end{table}

\begin{table}[htbp]
	\caption{Averaged estimation errors and standard errors of $\cD(\hat{\Rb},\Rb_{0})$, $ \cD(\hat{\Cb},\Cb_{0})$ for Setting B under $\cN(0,1), t_{5}, t_{3}$ distribution over 500 replications. ``IHR": Iterative Huber Regression. ``RMFM": robust matrix factor analysis method \citep{He2021MatrixFA}. ``PE": projection estimation method \citep{Yu2021Projected}. ``$\alpha$-PCA": $\alpha$-PCA with $\alpha=0$ \citep{fan2021}.} \label{table:2}
	\renewcommand{\arraystretch}{1.2}
	\centering
	\scalebox{0.9}{
		\begin{tabular}{cccccccc}
			\toprule[2pt]
			Evaluation &$T$ &$p_{1}$ &$p_{2}$ &IHR &RMFM &PE &$\alpha$-PCA($\alpha=0$)\\
			\cmidrule(r){1-8}
			\multicolumn{8}{c}{Normal Distribution}\\
			\multirow{5}{2cm}{\centering $\cD(\hat{\Rb},\Rb_{0})$}&
			20 &20 &\multirow{5}{1cm}{\centering 20}
			 &0.0938(0.0149)&0.0916(0.0146)&0.0916(0.0146)&0.0992(0.0172)\\
			&50&50&   &0.0579(0.0061)&0.0565(0.0059)&0.0565(0.0059)&0.0582(0.0064)\\
			&100&100& &0.0404(0.0033)&0.0395(0.0032)&0.0395(0.0032)&0.0400(0.0033)\\
			&150&150& &0.0330(0.0024)&0.0322(0.0023)&0.0322(0.0023)&0.0325(0.0024)\\
			&200&200& &0.0287(0.0022)&0.0280(0.0021)&0.0280(0.0021)&0.0282(0.0022)\\
			\cmidrule(r){2-8}
			\multirow{5}{2cm}{\centering $\cD(\hat{\Cb},\Cb_{0})$}&
			20 &20 &\multirow{5}{1cm}{\centering 20}
			 &0.0952(0.0171)&0.0929(0.0167)&0.0929(0.0168)&0.1013(0.0201)\\
			&50&50&   &0.0365(0.0051)&0.0355(0.0050)&0.0355(0.0050)&0.0389(0.0062)\\
			&100&100& &0.0178(0.0026)&0.0174(0.0026)&0.0174(0.0026)&0.0191(0.0032)\\
			&150&150& &0.0119(0.0016)&0.0116(0.0015)&0.0116(0.0015)&0.0127(0.0019)\\
			&200&200& &0.0089(0.0013)&0.0087(0.0013)&0.0087(0.0013)&0.0095(0.0016)\\
			\hline
			\multicolumn{8}{c}{$t_{5}$ Distribution}\\
			\multirow{5}{2cm}{\centering $\cD(\hat{\Rb},\Rb_{0})$}&
			20 &20 &\multirow{5}{1cm}{\centering 20}
			 &0.1086(0.0182)&0.1209(0.0203)&0.1213(0.0205)&0.1426(0.0305)\\
			&50&50&   &0.0663(0.0067)&0.0734(0.0076)&0.0735(0.0077)&0.0774(0.0086)\\
			&100&100& &0.0460(0.0038)&0.0509(0.0042)&0.0509(0.0042)&0.0522(0.0044)\\
			&150&150& &0.0376(0.0027)&0.0417(0.0029)&0.0417(0.0029)&0.0424(0.0031)\\
			&200&200& &0.0326(0.0025)&0.0360(0.0029)&0.0361(0.0029)&0.0365(0.0029)\\
			\cmidrule(r){2-8}
			\multirow{5}{2cm}{\centering $\cD(\hat{\Cb},\Cb_{0})$}&
			20 &20 &\multirow{5}{1cm}{\centering 20}
			 &0.1104(0.0196)&0.1227(0.0220)&0.1231(0.0223)&0.1451(0.0344)\\
			&50&50&   &0.0412(0.0060)&0.0460(0.0068)&0.0460(0.0069)&0.0551(0.0114)\\
			&100&100& &0.0206(0.0030)&0.0228(0.0033)&0.0228(0.0033)&0.0273(0.0051)\\
			&150&150& &0.0135(0.0018)&0.0150(0.0020)&0.0150(0.0020)&0.0179(0.0031)\\
			&200&200& &0.0103(0.0015)&0.0113(0.0016)&0.0113(0.0016)&0.0137(0.0029)\\
			\hline
			\multicolumn{8}{c}{$t_{3}$ Distribution}\\
			\multirow{5}{2cm}{\centering $\cD(\hat{\Rb},\Rb_{0})$}&
			20 &20 &\multirow{5}{1cm}{\centering 20}
			 &0.1213(0.0206)&0.2463(0.1496)&0.2597(0.1545)&0.3263(0.1460)\\
			&50&50&   &0.0724(0.0076)&0.1290(0.1112)&0.1385(0.1193)&0.1663(0.1228)\\
			&100&100& &0.0505(0.0041)&0.0795(0.0703)&0.0864(0.0845)&0.0961(0.0854)\\
			&150&150& &0.0412(0.0029)&0.0677(0.0757)&0.0721(0.0860)&0.0768(0.0846)\\
			&200&200& &0.0356(0.0028)&0.0543(0.0573)&0.0555(0.0577)&0.0585(0.0577)\\
			\cmidrule(r){2-8}
			\multirow{5}{2cm}{\centering $\cD(\hat{\Cb},\Cb_{0})$}&
			20 &20 &\multirow{5}{1cm}{\centering 20}
			 &0.1208(0.0221)&0.2466(0.1491)&0.2589(0.1532)&0.3316(0.1444)\\
			&50&50&   &0.0457(0.0071)&0.0964(0.1197)&0.1039(0.1272)&0.1651(0.1287)\\
			&100&100& &0.0226(0.0032)&0.0435(0.0754)&0.0481(0.0875)&0.0973(0.0997)\\
			&150&150& &0.0146(0.0020)&0.0335(0.0823)&0.0362(0.0903)&0.0727(0.0969)\\
			&200&200& &0.0112(0.0016)&0.0214(0.0576)&0.0218(0.0580)&0.0534(0.0730)\\
			\bottomrule[2pt]
		\end{tabular}
	}
\end{table}

    \subsection{Estimating the loading spaces}\label{sec:4.2}
    In this section, we investigate the accuracy of the estimated loading spaces by different approaches and the pair of factor numbers are given as a priori. In detail,
    we compare the performances of our Iterative Huber Regression (IHR) method with the PE by \cite{Yu2021Projected}, the RMFA by \cite{He2021MatrixFA} and the $\alpha$-PCA by \cite{fan2021}. For the implementation of  IHR,  we need to specify the initial estimators for $\Rb_{0}$ and $\Cb_{0}$, and we adopt the random projections, i.e., the initial estimators $\hat{\Rb}^{(0)},\hat{\Cb}^{(0)}$ are  randomly chosen and all their entries are sampled from a standard normal distribution.  We consider the following two settings:

    \vspace{0.5em}

	Setting A: $p_{1}=20, T=p_{2} \in \{20,50,100,150,200\}, \phi=0.1, \psi=0.1.$
	
    \vspace{0.5em}

	Setting B: $p_{2}=20, T=p_{1} \in \{20,50,100,150,200\}, \phi=0.1, \psi=0.1.$

    \vspace{0.5em}
	
	To measure the performances of various methods in terms of estimating the loading spaces, we adopt a metric quantifying the distance between two linear spaces which was also utilized in \cite{Yu2021Projected}, \cite{He2021MatrixFA}. For two column-wise orthogonal matrices $(\bQ_1)_{p\times q_1}$ and $(\bQ_2)_{p\times q_2}$, we define
	\[
	 \mathcal{D}(\bQ_1,\bQ_2)=\bigg(1-\frac{1}{\max{(q_1,q_2)}}\mbox{Tr}\Big(\bQ_1\bQ_1^{\top}\bQ_2\bQ_2^{\top}\Big)\bigg)^{1/2}.
	\]
	By the definition of $\mathcal{D}(\bQ_1,\bQ_2)$, it holds that $0\leq \mathcal{D}(\bQ_1,\bQ_2)\leq 1$, and in essence $\mathcal{D}(\bQ_1,\bQ_2)$ measures the distance between the column spaces spanned by  $\bQ_1$ and $\bQ_2$, i.e., $\text{span}(\bQ_1)$ and $\text{span}(\bQ_2)$. In particular, $\text{span}(\bQ_1)$ and $\text{span}(\bQ_2)$  are the same when $\mathcal{D}(\bQ_1,\bQ_2)=0$, while  $\text{span}(\bQ_1)$ and $\text{span}(\bQ_2)$   are orthogonal when $\mathcal{D}(\bQ_1,\bQ_2)=1$.  The Gram-Schmidt orthogonalization can be used to make $\bQ_1$ and $\bQ_2$  column-orthogonal matrices if they are not.
	
	Table \ref{table:1} and Table \ref{table:2} show the averaged errors with standard errors in parentheses for normal distribution and heavy-tailed  $t_{5}$ and $t_{3}$, under Settings A and B respectively . From these two tables, we can draw the following conclusions. Firstly, in the case that  the idiosyncratic errors are from normal distribution, all the methods benefit from large dimensions and sample sizes, and the four methods preform comparably well. As we know, Huber loss acts as the combination of least squares and least absolute, and when  errors come from a light-tailed distribution such as  normal distribution, the absolute value of the sampled errors will be less than $\tau$, hence, the methods based on the Huber loss (IHR, RMFM) and the method based on least squares loss (PE) performs almost the same.  \cite{He2021MatrixFA} provided a least square interpretation of the PE method by \cite{Yu2021Projected}, and that's why PE always shows the lowest estimation errors and standard errors compared with the other methods in the normal case. Secondly, when the entries of the idiosyncratic error matrices are from the $t_{5}$ or $t_{3}$ distribution, the picture is completely different. Although all the methods benefit from large dimensions, the RMFA and IHR  methods show great advantage over both PE and $\alpha$-PCA methods in all settings. It is also worth noting that the IHR method also outperforms the RMFA by a large margin  when data are heavy-tailed, especially for the $t_3$ case.
    In a word, the IHR performs robustly and much better than other methods when  the idiosyncratic errors are heavy-tailed and performs almost the same with other mathods when the data have light tails. As a result, the IHR method can be used as a safe replacement of the RMFA, $\alpha$-PCA and PE in real applications.

\begin{table}[htbp]
	\caption{The frequencies of exact estimation and underestimation of the numbers of factors under Setting A and B over 100 replications. ``IHR-RM": the proposed rank-minimization based on IHR. ``IHR-ER": the proposed eigenvalue-ratio based on IHR. ``Rit-ER": the robust iterative eigenvalue-ratio based method \citep{He2021MatrixFA}. ``IterER": iterative eigenvalue-ratio based method \citep{Yu2021Projected}. ``$\alpha$-PCA-ER": $\alpha$-PCA based eigenvalue-ratio method with $\alpha=0$ \citep{fan2021}.} \label{table:3}
	\renewcommand{\arraystretch}{1.2}
	\centering
	\scalebox{1}{
		\begin{tabular}{ccccccc}
			\toprule[2pt]
			Distribution &$T$ &IHR-RM &IHR-ER &Rit-ER &IterER &$\alpha$-PCA-ER  \\
			\cmidrule(r){1-7}
			\multicolumn{7}{c}{Setting A: $p_{1}=20, p_{2}=T$} \\
			\multirow{5}{1cm}{\centering Normal}
			&20  &0.9860(0.0140) &0.9560(0.0440) &0.9940(0.0060) &0.9940(0.0060) &0.8340(0.1660)\\
			&50  &1.0000(0.0000) &1.0000(0.0000) &1.0000(0.0000) &1.0000(0.0000) &0.9840(0.0160)\\
			&100 &1.0000(0.0000) &1.0000(0.0000) &1.0000(0.0000) &1.0000(0.0000) &0.9860(0.0140)\\
			&150 &1.0000(0.0000) &1.0000(0.0000) &1.0000(0.0000) &1.0000(0.0000) &0.9900(0.0100)\\
			&200 &1.0000(0.0000) &1.0000(0.0000) &1.0000(0.0000) &1.0000(0.0000) &0.9920(0.0080)\\
			\hline
			\multirow{5}{1cm}{\centering $t_{5}$}
			&20  &0.8800(0.1200) &0.9060(0.0940) &0.9700(0.0300) &0.9680(0.0320) &0.5820(0.4180)\\
			&50  &1.0000(0.0000) &0.9980(0.0020) &1.0000(0.0000) &1.0000(0.0000) &0.9420(0.0580)\\
			&100 &1.0000(0.0000) &1.0000(0.0000) &1.0000(0.0000) &1.0000(0.0000) &0.9660(0.0340)\\
			&150 &1.0000(0.0000) &1.0000(0.0000) &1.0000(0.0000) &1.0000(0.0000) &0.9740(0.0260)\\
			&200 &1.0000(0.0000) &1.0000(0.0000) &1.0000(0.0000) &1.0000(0.0000) &0.9800(0.0200)\\
			\hline
			\multirow{5}{1cm}{\centering $t_{3}$}
			&20  &0.4600(0.5400) &0.7240(0.2760) &0.5440(0.4560) &0.5180(0.4820) &0.1540(0.8460)\\
			&50  &1.0000(0.0000) &0.9960(0.0040) &0.8420(0.1580) &0.7480(0.2520) &0.5460(0.4540)\\
			&100 &1.0000(0.0000) &1.0000(0.0000) &0.9380(0.0620) &0.8200(0.1800) &0.6920(0.3080)\\
			&150 &1.0000(0.0000) &1.0000(0.0000) &0.9560(0.0440) &0.8740(0.1260) &0.7980(0.2020)\\
			&200 &1.0000(0.0000) &1.0000(0.0000) &0.9840(0.0160) &0.8960(0.1040) &0.8220(0.1780)\\
			\hline
			\multicolumn{7}{c}{Setting B: $p_{2}=20, p_{1}=T$} \\
			\multirow{5}{1cm}{\centering Normal}
			&20  &0.9860(0.0140) &0.9560(0.0440) &0.9940(0.0060) &0.9940(0.0060) &0.8340(0.1660)\\
			&50  &1.0000(0.0000) &1.0000(0.0000) &1.0000(0.0000) &1.0000(0.0000) &0.9760(0.0240)\\
			&100 &1.0000(0.0000) &1.0000(0.0000) &1.0000(0.0000) &1.0000(0.0000) &0.9800(0.0200)\\
			&150 &1.0000(0.0000) &1.0000(0.0000) &1.0000(0.0000) &1.0000(0.0000) &0.9900(0.0100)\\
			&200 &1.0000(0.0000) &1.0000(0.0000) &1.0000(0.0000) &1.0000(0.0000) &0.9920(0.0080)\\
			\hline
			\multirow{5}{1cm}{\centering $t_{5}$}
			&20  &0.8800(0.1200) &0.9060(0.0940) &0.9700(0.0300) &0.9680(0.0320) &0.5820(0.4180)\\
			&50  &1.0000(0.0000) &1.0000(0.0000) &1.0000(0.0000) &0.9980(0.0020) &0.9480(0.0520)\\
			&100 &1.0000(0.0000) &1.0000(0.0000) &1.0000(0.0000) &1.0000(0.0000) &0.9600(0.0400)\\
			&150 &1.0000(0.0000) &1.0000(0.0000) &1.0000(0.0000) &1.0000(0.0000) &0.9800(0.0200)\\
			&200 &1.0000(0.0000) &1.0000(0.0000) &1.0000(0.0000) &0.9980(0.0020) &0.9740(0.0260)\\
			\hline
			\multirow{5}{1cm}{\centering $t_{3}$}
			&20  &0.4600(0.5400) &0.7240(0.2760) &0.5440(0.4560) &0.5180(0.4820) &0.1540(0.8460)\\
			&50  &1.0000(0.0000) &0.9980(0.0020) &0.8040(0.1960) &0.7380(0.2620) &0.5340(0.4660)\\
			&100 &1.0000(0.0000) &1.0000(0.0000) &0.8920(0.1080) &0.8080(0.1920) &0.6940(0.3060)\\
			&150 &1.0000(0.0000) &1.0000(0.0000) &0.9220(0.0780) &0.8600(0.1400) &0.8000(0.2000)\\
			&200 &1.0000(0.0000) &1.0000(0.0000) &0.9560(0.0440) &0.9000(0.1000) &0.8120(0.1880)\\
			\bottomrule[2pt]
		\end{tabular}
	}
\end{table}

\subsection{Estimating the numbers of factors}\label{sec:4.3}
Determining the number of factors is the first step to do factor analysis. In this section, we verify the validity of our rank-minimization and eigenvalue-ratio methods (denoted as IHR-RM and IHR-ER respectively) for determining the pair of factor numbers. We compare the proposed methods with the state-of-the-art ones including Iter-ER by  \cite{Yu2021Projected}, Rit-ER by \cite{He2021MatrixFA} and $\alpha$-PCA-ER by \citep{fan2021}. For rank minimization, similar to the choice by \cite{Chen2021Quantile}, let $P_{1}=\hat{\sigma}_{1,1}D^{-2/3},P_{2}=\hat{\sigma}_{2,1}D^{-2/3}$ in (\ref{equ:rankfac}). The maximum factor numbers $m_{1}, m_{2}$ are set as 6.

Table \ref{table:3} presents the frequencies  of exact estimation ($(\hat{k}_{1},\hat{k}_{2})=(3,3)$) and underestimation over 500 replications under Setting A and Setting B by different methods. Under the normal case, we can see that the  IHR-RM, IHR-ER, Rit-ER, IterER have higher estimation accuracy and lower underestimation rates compared with  $\alpha$-PCA-ER, and the methods perform well even for small $T=p_{1}=p_{2}=20$. When $T=p_{2}=50$, the IHR-RM, IHR-ER, Rit-ER, IterER can always estimate the number of factors correctly over 500 replications. With the increase of dimensions, the accuracy of all methods increases gradually. As the idiosyncratic errors become heavy-tailed, such as $t_{3}$ distribution, although all the methods perform worse in the case, the IHR-ER has higher estimation accuracy than the other methods when $T=p_{1}=p_{2}=20$. When  $T=50$, both the IHR-RM and IHR-ER outperforms the other methods. In other words, the proposed two methods perform robustly and have higher estimation accuracy especially for heavy-tailed idiosyncratic errors, and as the sample size $T$ increases, the accuracy rate of our methods IHR-RM, IHR-ER can gradually converge to 1.

\begin{figure}[!h]
	 \centerline{\includegraphics[width=16cm,height=12cm]{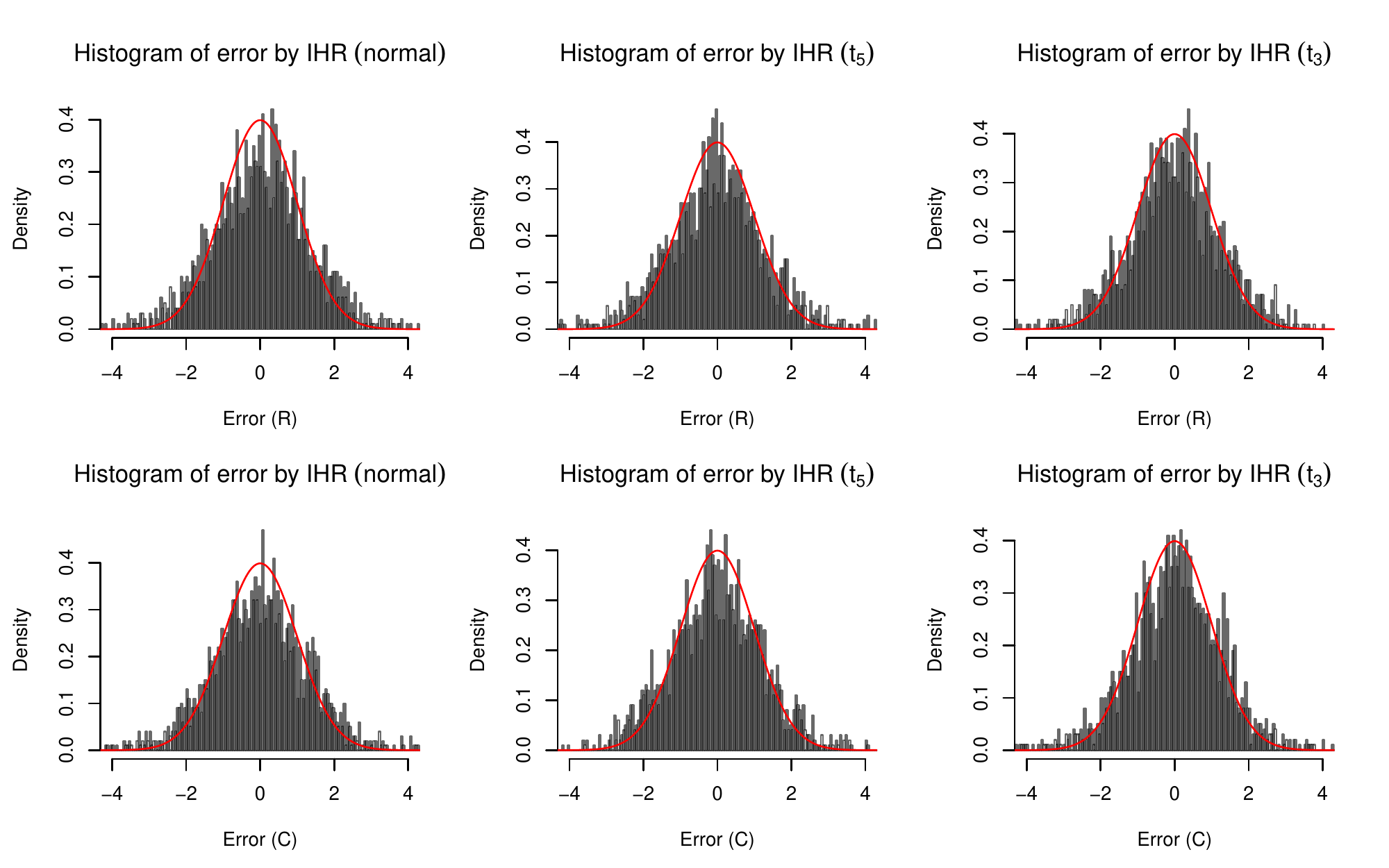}}
	\caption{Histograms of estimation errors for $r_{i,1}$ and $c_{j,1}$ after normalization over 2000 replications. The histograms in top panel are the estimation errors for $r_{i,1}$ under normal, $t_{5}$ and $t_{3}$ with $T=p_{2}=10,p_{1}=100$. The  histograms in the bottom panel are the estimation errors for $c_{j,1}$ under normal, $t_{5}$ and $t_{3}$ with $T=p_{1}=10,p_{2}=100$, respectively. The red real line plots the probability density function of the standard normal distribution.}
	\label{fig:2}
\end{figure}

\subsection{Verifying the asymptotic normality}\label{sec:4.4}
In this section, we verify the asymptotic normality of the estimators for factor loadings  derived in Theorem \ref{Theorem 4} by numerical studies. We adopt the same data generation mechanism introduced in Section \ref{sec:4.1}, and we normalize $\Rb_{0}$, $\Cb_{0}$ and $\Fb_{0t}$ by (\ref{Normalized R}), (\ref{Normalized C}), (\ref{Normalized F}), respectively, such that the identification conditions are satisfied. The idiosyncratic errors $\Eb_{t}$ are generated with  $\psi=0$ such that $e_{t,ij}$ are independent across $t,i,j$ and $e_{t,ij}$ are {\it i.i.d.} sampled from $\cN(0,1), t_{5}$ or $t_{3}$. The parameter $\tau$ is fixed and we set it to be $1.345\sigma$ such that when the error distribution is normal the Huber estimator is $95\%$ efficient. In practice, the parameter $\sigma$ is estimated by $\hat{s} = 1.483 \cdot \text{median}\{|\hat{e}_{t,ij}|\}_{t,i,j}$, and $\hat{e}_{t,ij}$
is the  residuals estimated by the Projection Estimation method by \cite{Yu2021Projected}.
In fact, we tried different $\tau$ and the simulation results show that  different choices of the parameter $\tau$ would lead to the same conclusions in the following (see more simulation results in the supplement for different $\tau$).

\begin{figure}[!h]
	 \centerline{\includegraphics[width=16cm,height=12cm]{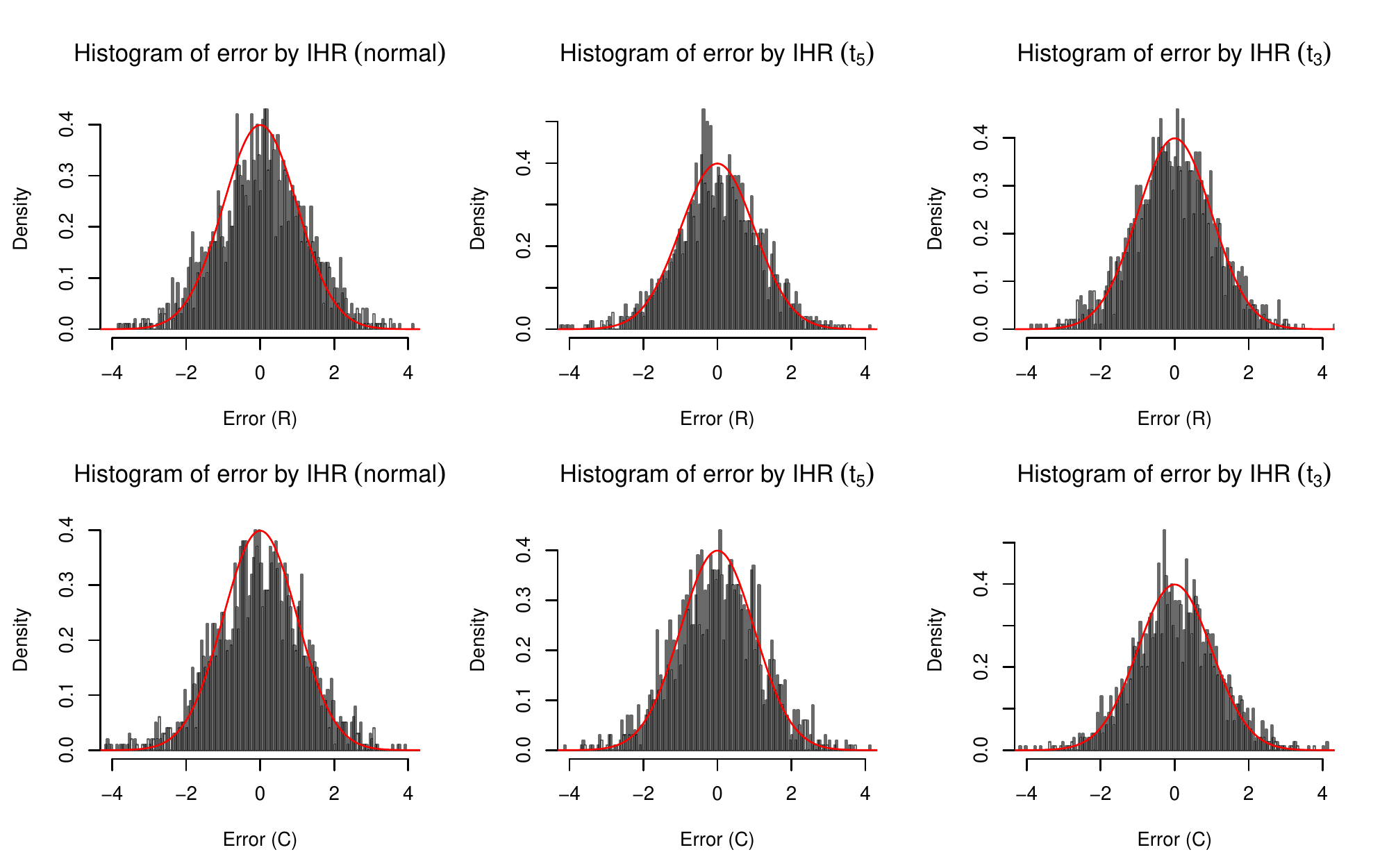}}
	\caption{Histograms of estimation errors for $r_{i,1}$ and $c_{j,1}$ after normalization over 2000 replications. The histograms in top panel are the estimation errors for $r_{i,1}$ under normal, $t_{5}$ and $t_{3}$ with $T=p_{2}=50,p_{1}=500$. In the bottom panel are the estimation errors for $c_{j,1}$ under normal, $t_{5}$ and $t_{3}$ $T=p_{1}=50,p_{2}=500$, respectively. The red real curve plots the probability density function of the standard normal distribution.}
	\label{fig:3}
\end{figure}

Figure \ref{fig:2} and Figure \ref{fig:3} show the histograms of the first coordinates of $\sqrt{Tp_{2}}\hat{\bSigma}_{Tp_{2},i}^{-1/2}\hat{\bPhi}_{i}(\hat{\br}_{i}-\hat{\Hb}_{1}\br_{0i})$ at $i=\lfloor p_{1}/2 \rfloor$ from 2000 repetitions when $T=p_{2}, (p_{1},p_{2}) \in \{(100,10),(500,50)\}$, and the first coordinates of $\sqrt{Tp_{1}}\hat{\bSigma}_{Tp_{1},j}^{-1/2}\hat{\bPsi}_{j}(\hat{\bc}_{j}-\hat{\Hb}_{2}\bc_{0j})$ at $j=\lfloor p_{2}/2 \rfloor$ from 2000 repetitions when $T=p_{1}, (p_{1},p_{2}) \in \{(10,100),(50,500)\}$, where $\hat{\bSigma}_{Tp_{2},i}, \hat{\bPhi}_{i}$ and $\hat{\bSigma}_{Tp_{1},j}, \hat{\bPsi}_{j}$ are the estimators introduced in Section 3.3. The results show that the asymptotic distributions in Theorem \ref{Theorem 4} provide a good fit in the finite-sample case, even for $T=p_{2}=10$ or $T=p_{1}=10$,  under both normal and heavy-tailed $t_{5}$ or $t_{3}$ distributions of the idiosyncratic errors.

\section{Real Example}
In this section, we illustrate the empirical usefulness of  the proposed methods by two real examples. The first one is a $10\times 10$ Fama-French portfolios and the second one is a macroeconomic indices dataset of China which contains 81 macroeconomic indices across 30 provinces $(81\times 30)$. Clearly, the first dataset is of relatively low dimensions compared with the second one. Similar as in the simulation studies, we also take the PE, RMFM and $\alpha$-PCA for comparison in these two examples.
\subsection{Fama-French Portfolios}
\begin{figure}[htbp]
	\centering
	\subfigure[Histogram of the sample kurtosis for 100 portfolios.]{\includegraphics[height=7cm,width=7cm]{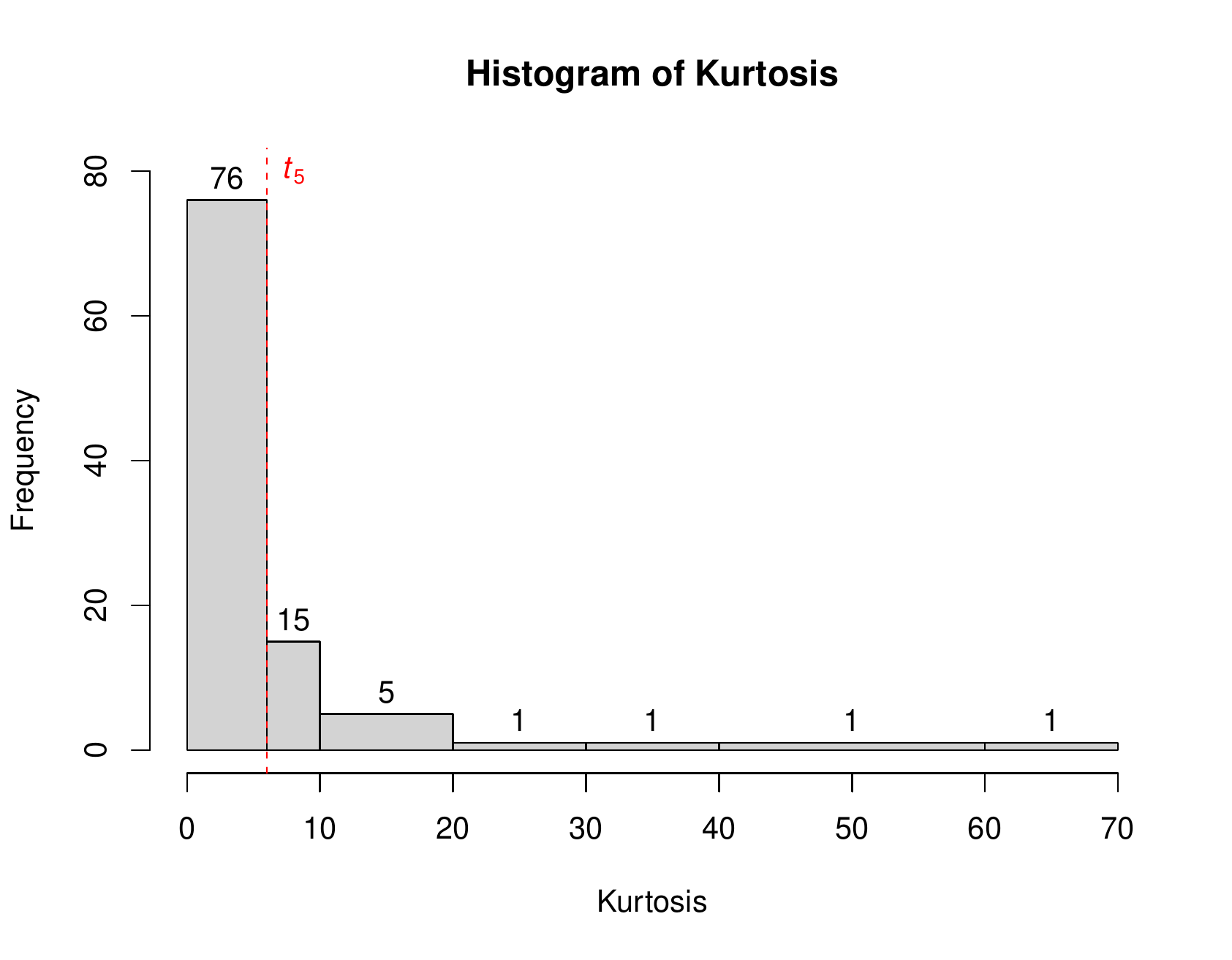}}
	\subfigure[Histogram of the sample kurtosis for multinational macroeconomic indices of China.]{\includegraphics[height=7cm,width=8cm]{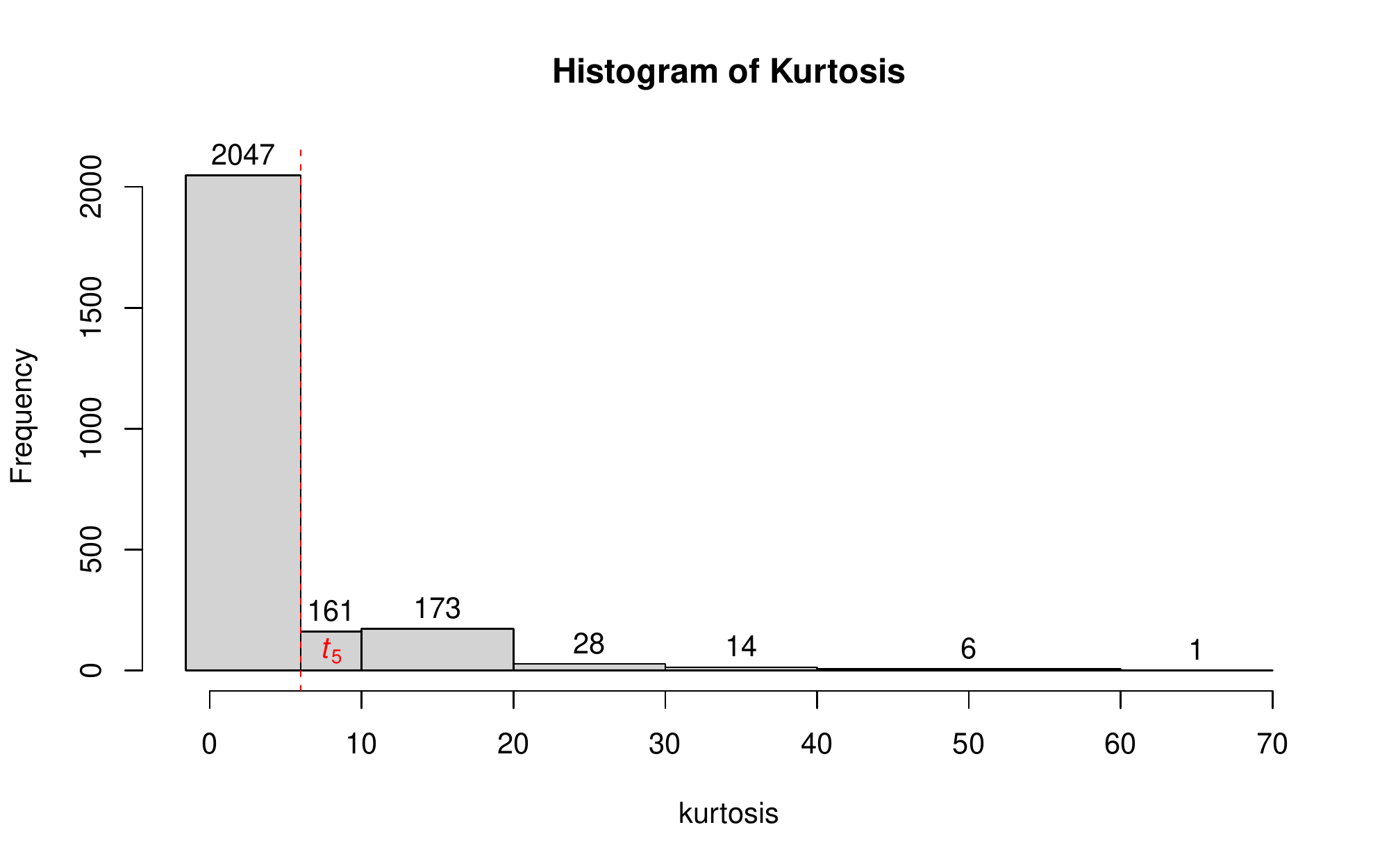}}
	\caption{Histogram of the sample kurtosis and the red dashed line is the theoretical kurtosis of $t_5$ distribution.}
	\label{fig:4}
\end{figure}

The portfolios are the intersections of 10 level portfolios formed on size (S1-S10) and 10 level portfolios formed on operating profitability (OP1-OP10). It contains the monthly return series of 100 portfolios and we use the data from 2003-03 to 2023-02 in the study, covering 240 months without missing values. The dataset is  open access provided by Kenneth R. French, which can be downloaded from the website \url{http://mba.tuck.dartmouth.edu/pages/faculty/ken.french/data_library.html}. In addition,  the excess market returns is also provided by this website. We follow the same preprocessing steps of {\cite{wang2019factor}} and \cite{Yu2021Projected} by subtracting the excess market returns and standardizing each of the portfolio return series. The augmented Dickey-Fuller test rejects the null hypotheses for all the series, which indicates the stationarity of all series. The histogram of the sample kurtosis for the 100 portfolios are reported in Figure \ref{fig:4} (a), which indicates the heavy-tailedness.

For the factor numbers, the Rit-ER by \cite{He2021MatrixFA}, Iter-ER by \cite{Yu2021Projected} and $\alpha$-PCA-ER by \cite{fan2021} all suggest $(k_1,k_2)=(1,1)$ while the proposed IHR-RM suggests $(k_1,k_2)=(2,1)$. For better illustration, we take $(k_1,k_2)=(2,2)$ for all methods, and the estimated row and column loading matrices after varimax rotation and scaling are reported in Table \ref{table:4}. From the perspective of size, the small size portfolios load heavily on the first factor while the large size portfolios load mainly on the second factor. From the perspective of operating profitability, the small OP portfolios load heavily on the second factor while the large OP portfolios load mainly on the first factor.

We next use a rolling-validation procedure to compare the performance of the proposed IHR against the $\alpha$-PCA, PE and RMFA. For each year $t$ form 2013 to 2023, we repeatedly use the $n$ (bandwidth) years observations before $t$ to fit the matrix-variate factor model and estimate the two loading matrices. The loadings are then used to estimate the factors and corresponding residuals of the
12 months in the current year. In specific, let $\hat{\Yb}_{t+i}$ be the estimated price matrix of month $i$ in year $t$, $\Yb_{t+i}$ be the corresponding observed price matrix and  $\bar{\Yb}_t$ be the mean price matrix, we employ the following metric to compare different methods:
$$\text{MSE}_t=\frac{1}{12 \times p_{1} \times p_{2}}\sum_{i=1}^{12}||\hat{\Yb}_{t+i}-\Yb_{t+i}||_{\text{F}}^2,\quad \rho_t=\frac{\sum_{i=1}^{12}||\hat{\Yb}_{t+i}-\Yb_{t+i}||_{\text{F}}^2}{\sum_{i=1}^{12}||\hat{\Yb}_{t+i}-\bar{\Yb}_{t}||_{\text{F}}^2},$$
which are the mean squared pricing error and unexplained proportion of total variances, respectively. In the rolling-validation procedure,
the variation of the loading space is measured by $v_t:=\cD(\hat{\Cb}_t\otimes\hat{\Rb}_t,\hat{\Cb}_{t-1}\otimes\hat{\Rb}_{t-1})$. 
We tried $n\in\cbr{5,10}$ and the numbers of factors $(k_1,k_2)\in\cbr{(1,1),(1,2),(2,1),(2,2)}$ and we report the means of $\text{MSE}_t$, $\rho_t$ and $v_t$ in Table \ref{table:5}, which implies that IHR outperforms the others in almost all settings.

\begin{table}[!h]
	\caption{Loading matrices for Fama-French portfolios after varimax rotation and scaling by 30.} \label{table:4}
	\renewcommand{\arraystretch}{1}
	\centering
	\scalebox{0.9}{
		\begin{tabular}{cccccccccccc}
			\toprule[2pt]
			\multicolumn{12}{c}{Size}\\
			\cmidrule(r){1-12}
			Method &Factor &S1 &S2 &S3 &S4 &S5 &S6 &S7 &S8 &S9 &S10  \\
			\cmidrule(r){1-12}
			\multirow{2}{*}{IHR}
			&1&\cellcolor {Lavender}47&\cellcolor {Lavender}51&\cellcolor {Lavender}43&\cellcolor {Lavender}35&\cellcolor {Lavender}26&13&8&-2&-9&-12\\
			&2&17&12&-5&-13&-25&\cellcolor {Lavender}-38&\cellcolor {Lavender}\cellcolor {Lavender}-43&\cellcolor {Lavender}-46&\cellcolor {Lavender}-43&\cellcolor {Lavender}-24\\
			\cmidrule(r){1-12}
			\multirow{2}{*}{RMFM}
			&1&\cellcolor {Lavender}43&\cellcolor {Lavender}50&\cellcolor {Lavender}42&\cellcolor {Lavender}36&\cellcolor {Lavender}28&16&9&0&-10&-16\\
			&2&13&12&-5&-12&-23&\cellcolor {Lavender}-34&\cellcolor {Lavender}-41&\cellcolor {Lavender}-44&\cellcolor {Lavender}-45&\cellcolor {Lavender}-33\\
			\cmidrule(r){1-12}
			\multirow{2}{*}{PE}
			&1&\cellcolor {Lavender}-45&\cellcolor {Lavender}-51&\cellcolor {Lavender}-42&\cellcolor {Lavender}-35&\cellcolor {Lavender}-27&-16&-10&0&10&16\\
			&2&14&12&-6&-12&-24&\cellcolor {Lavender}-32&\cellcolor {Lavender}-41&\cellcolor {Lavender}-43&\cellcolor {Lavender}-47&\cellcolor {Lavender}-32\\
			\cmidrule(r){1-12}
			\multirow{2}{*}{$\alpha$-PCA}
			&1&\cellcolor {Lavender}34&\cellcolor {Lavender}40&\cellcolor {Lavender}40&\cellcolor {Lavender}37&\cellcolor {Lavender}34&\cellcolor {Lavender}28&\cellcolor {Lavender}24&16&-3&-21\\
			&2&7&6&0&-3&-10&-16&\cellcolor {Lavender}-24&\cellcolor {Lavender}-29&\cellcolor {Lavender}-55&\cellcolor {Lavender}-65\\
			\cmidrule(r){1-12}
			\multicolumn{12}{c}{Operating Profitability}\\
			\cmidrule(r){1-12}
			Method &Factor &BE1 &BE2 &BE3 &BE4 &BE5 &BE6 &BE7 &BE8 &BE9 &BE10 \\
			\cmidrule(r){1-12}
			\multirow{2}{*}{IHR}
			&1&-14&13&\cellcolor {Lavender}30&\cellcolor {Lavender}35&\cellcolor {Lavender}35&\cellcolor {Lavender}37&\cellcolor {Lavender}34&\cellcolor {Lavender}31&\cellcolor {Lavender}32&\cellcolor {Lavender}29\\
			&2&\cellcolor {Lavender}83&\cellcolor {Lavender}42&13&3&0&-6&-4&5&2&8\\
			\cmidrule(r){1-12}
			\multirow{2}{*}{RMRM}
			&1&-13&11&\cellcolor {Lavender}28&\cellcolor {Lavender}34&\cellcolor {Lavender}33&\cellcolor {Lavender}39&\cellcolor {Lavender}34&\cellcolor {Lavender}32&\cellcolor {Lavender}32&\cellcolor {Lavender}31\\
			&2&\cellcolor {Lavender}81&\cellcolor {Lavender}46&16&6&3&-9&-2&2&1&3\\
			\cmidrule(r){1-12}
			\multirow{2}{*}{PE}
			&1&-12&12&\cellcolor {Lavender}29&\cellcolor {Lavender}34&\cellcolor {Lavender}32&\cellcolor {Lavender}39&\cellcolor {Lavender}33&\cellcolor {Lavender}33&\cellcolor {Lavender}33&\cellcolor {Lavender}31\\
			&2&\cellcolor {Lavender}81&\cellcolor {Lavender}47&13&3&3&-10&-2&3&2&4\\
			\cmidrule(r){1-12}
			\multirow{2}{*}{$\alpha$-PCA}
			&1&12&-7&\cellcolor {Lavender}-26&\cellcolor {Lavender}-32&\cellcolor {Lavender}-30&\cellcolor {Lavender}-41&\cellcolor {Lavender}-37&\cellcolor {Lavender}-33&\cellcolor {Lavender}-34&\cellcolor {Lavender}-32\\
			&2&\cellcolor {Lavender}74&\cellcolor {Lavender}53&20&6&11&-11&-8&3&0&3\\
			\bottomrule[2pt]
		\end{tabular}
	}
\end{table}

\begin{table}[htbp]
	\caption{Rolling validation for the Fama-French portfolios. $12n$ is the sample size of the training set. $(k_1,k_2)$ is the number of factors. $\overline{\text{MSE}}$, $\bar{\rho}$, $\bar{v}$ are the mean pricing error, mean unexplained proportion of total variances and mean variation of the estimated loading space.}\label{table:5}
	\renewcommand{\arraystretch}{1.2}
	\centering
	\scalebox{0.8}{
		\begin{tabular}{cccccccccccccc}
			\toprule[2pt]
			 \multirow{2}{*}{$n$}&\multirow{2}{*}{$(k_1,k_2)$}&\multicolumn{4}{c}{$\overline{\text{MSE}}$}&\multicolumn{4}{c}{$\bar{\rho}$}&\multicolumn{4}{c}{$\bar{v}$}\\
			 \cmidrule(r){3-6}\cmidrule(r){7-10}\cmidrule(r){11-14}
			 &&$\alpha$-PCA&\text{PE}&\text{RMFM}&\text{IHR}&$\alpha$-PCA&\text{PE}&\text{RMFM}&\text{IHR}&$\alpha$-PCA&\text{PE}&\text{RMFM}&\text{IHR}\\
			\hline
			5&(1,1)&0.6978 &0.6968 &0.6939 &\textbf{0.6928} &0.7271 &0.7239 &0.7202 &\textbf{0.7182} &0.0214 &0.0225 &0.0208 &\textbf{0.0204}\\
			10&(1,1)&0.7045 &0.7012 &0.6959 &\textbf{0.6949} &0.7348 &0.7290 &0.7215 &\textbf{0.7196} &0.0108 &0.0112 &0.0108 &\textbf{0.0106}\\
			\hline
			5&(1,2)&0.6401 &0.6329 &0.6299 &\textbf{0.6291} &0.6695 &0.6599 &0.6560 &\textbf{0.6543} &0.0581 &0.0386 &\textbf{0.0333} &0.0338\\
			10&(1,2)&0.6486 &0.6356 &0.6310 &\textbf{0.6290} &0.6821 &0.6636 &0.6570 &\textbf{0.6534} &0.0334 &0.0249 &\textbf{0.0219} &0.0244\\
			\hline
			5&(2,1)&0.6523 &0.6437 &\textbf{0.6416} &0.6426 &0.6832 &0.6719 &\textbf{0.6694} &0.6699 &0.0502 &0.0625 &\textbf{0.0485} &0.0546\\
			10&(2,1)&0.6605 &0.6444 &0.6430 &\textbf{0.6422} &0.6930 &0.6706 &0.6685 &\textbf{0.6675} &0.0238 &0.0221 &\textbf{0.0206} &0.0233\\
			\hline
			5&(2,2)&0.5830 &0.5698 &\textbf{0.5663} &0.5677 &0.6137 &0.5988 &\textbf{0.5947} &0.5951 &0.0759 &0.0691 &0.0545 &\textbf{0.0533}\\
			10&(2,2)&0.5917 &0.5738 &0.5715 &\textbf{0.5681} &0.6266 &0.6017 &0.5988 &\textbf{0.5937} &0.0410 &0.0343 &\textbf{0.0307} &0.0314\\
			\bottomrule[2pt]
		\end{tabular}
	}
\end{table}

\begin{figure}[!h]
	 \centerline{\includegraphics[width=16cm,height=9cm]{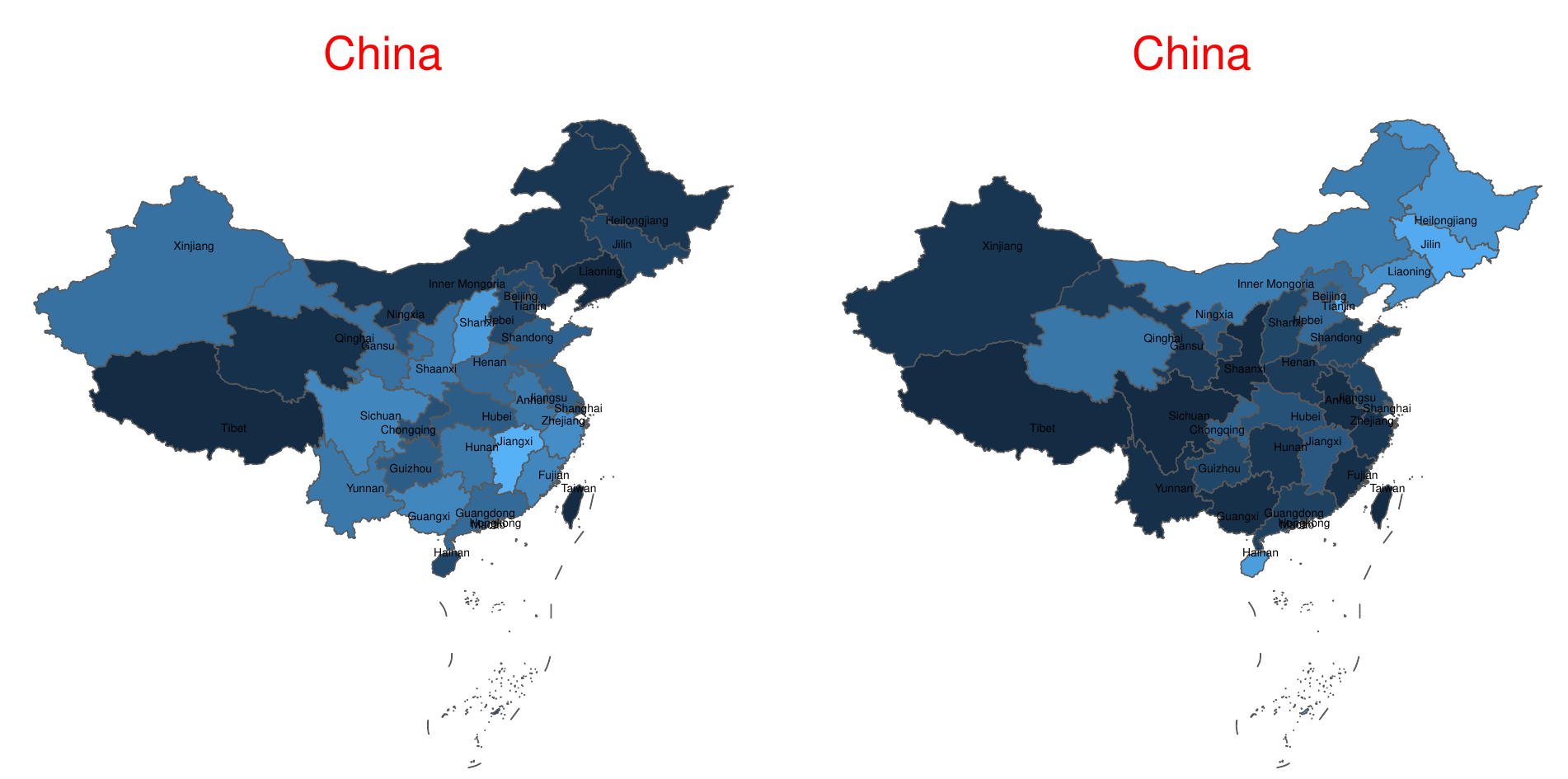}}
	\caption{Row (regions) loading matrix by IHR, varimax rotated and multiplied by 10. Left: the absolute value of the first row, right: the absolute value of the second row. The color becomes lighter as the absolute value increases.}
	\label{fig:5}
\end{figure}

\begin{figure}[htbp]
	\centering
	\subfigure[Heatmap by IHR.]{
		\includegraphics[width=7.5cm]{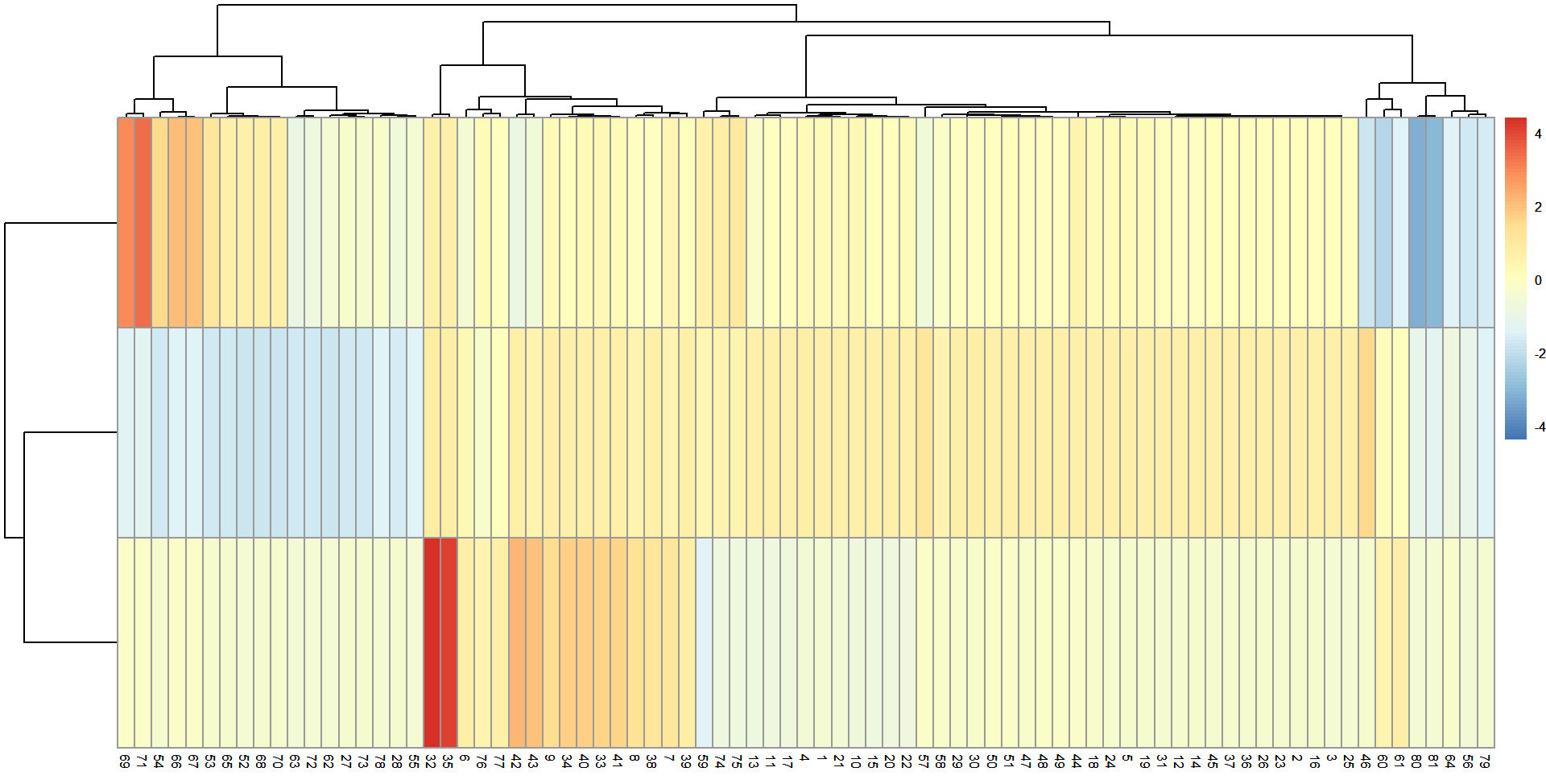}
	}
	\quad
	\subfigure[Hierarchical cluster by IHR.]{
		\includegraphics[width=8cm]{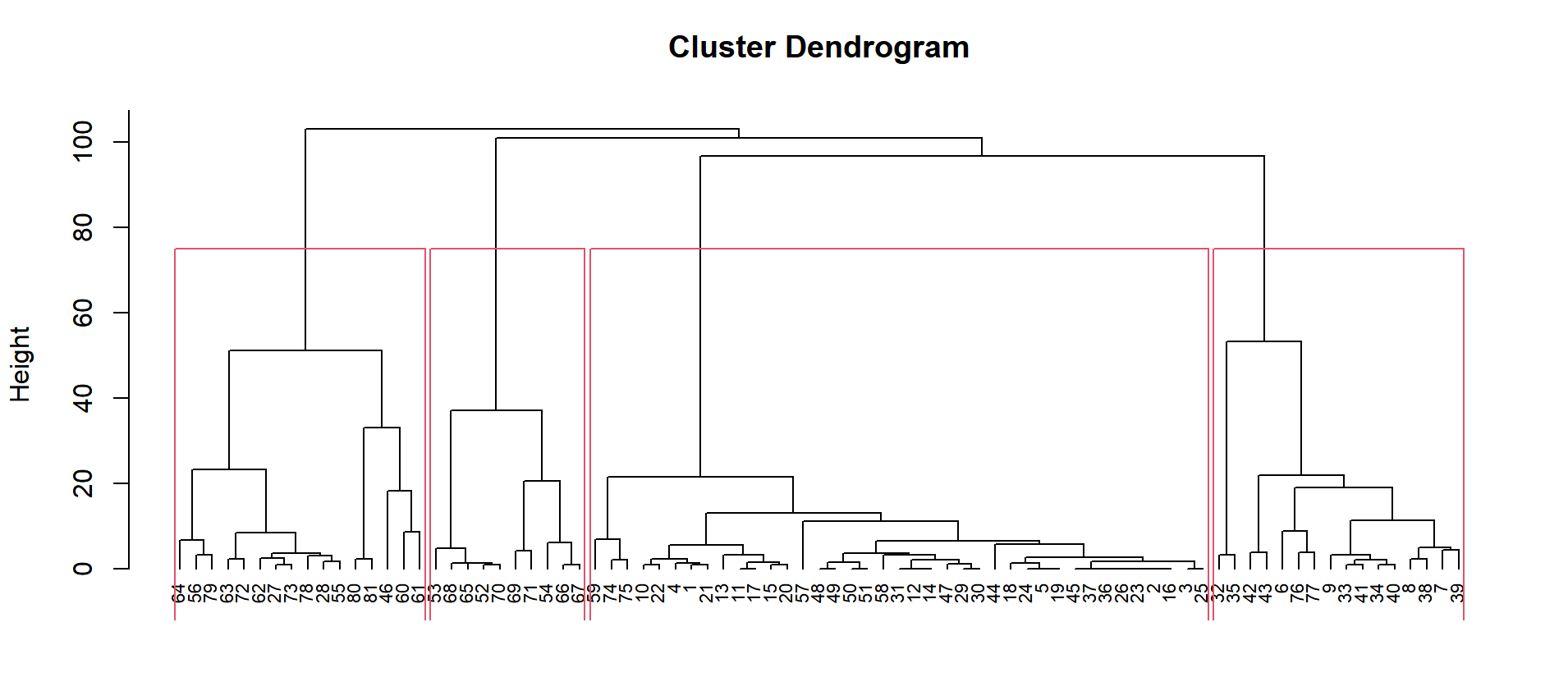}
	}
	\caption{The results for column loading matrix by IHR, varimax rotated and multiplied by 10.}
	\label{fig:6}
\end{figure}

\begin{table}[htbp]
	\caption{Rolling validation for the macroeconomic index dataset. $12n$ is the sample size of the training set. $(k_1,k_2)$ is the number of factors. $\overline{\text{MSE}}$, $\bar{\rho}$, $\bar{v}$ are the mean pricing error, mean unexplained proportion of total variances and mean variation of the estimated loading space.}\label{table:6}
	\renewcommand{\arraystretch}{1.2}
	\centering
	\scalebox{0.8}{
		\begin{tabular}{cccccccccccccc}
			\toprule[2pt]
			 \multirow{2}{*}{$n$}&\multirow{2}{*}{$(k_1,k_2)$}&\multicolumn{4}{c}{$\overline{\text{MSE}}$}&\multicolumn{4}{c}{$\bar{\rho}$}&\multicolumn{4}{c}{$\bar{v}$}\\
			 \cmidrule(r){3-6}\cmidrule(r){7-10}\cmidrule(r){11-14}
			 &&$\alpha$-PCA&\text{PE}&\text{RMFA}&\text{IHR}&$\alpha$-PCA&\text{PE}&\text{RMFA}&\text{IHR}&$\alpha$-PCA&\text{PE}&\text{RMFA}&\text{IHR}\\
			\hline
			1&(1,1)&0.4066 &0.4059 &0.4040 &\textbf{0.4018} &0.7111 &0.7099 &0.7066 &\textbf{0.7026} &0.0715 &0.0711 &0.0696 &\textbf{0.0629}\\
			2&(1,1)&0.4019 &0.4014 &0.4001 &\textbf{0.3989} &0.7027 &0.7019 &0.6995 &\textbf{0.6974} &0.0374 &0.0372 &0.0357 &\textbf{0.0319}\\
			3&(1,1)&0.4013 &0.4010 &0.4002 &\textbf{0.3987} &0.7016 &0.7012 &0.6997 &\textbf{0.6971} &0.0257 &0.0257 &0.0250 &\textbf{0.0222}\\
			\hline
			1&(1,3)&0.3621 &0.3608 &\textbf{0.3607} &0.3612 &0.6334 &0.6311 &\textbf{0.6309} &0.6316 &\textbf{0.2032} &0.2067 &0.2100 &0.2623\\
			2&(1,3)&0.3616 &0.3601 &0.3589 &\textbf{0.3563} &0.6323 &0.6294 &0.6274 &\textbf{0.6226} &0.1279 &0.1249 &\textbf{0.1238} &0.1607\\
			3&(1,3)&0.3614 &0.3595 &0.3587 &\textbf{0.3531} &0.6318 &0.6285 &0.6271 &\textbf{0.6169} &\textbf{0.0806} &0.0826 &0.0838 &0.1555\\
			\hline
			1&(1,4)&0.3536 &0.3525 &0.3522 &\textbf{0.3507} &0.6183 &0.6165 &0.6159 &\textbf{0.6135} &0.2575 &0.2435 &0.2390 &\textbf{0.2269}\\
			2&(1,4)&0.3516 &0.3483 &0.3477 &\textbf{0.3452} &0.6147 &0.6088 &0.6079 &\textbf{0.6035} &\textbf{0.1620} &0.2129 &0.2383 &0.2026\\
			3&(1,4)&0.3502 &0.3460 &0.3450 &\textbf{0.3428} &0.6124 &0.6046 &0.6028 &\textbf{0.5990} &0.1329 &0.1453 &0.1368 &\textbf{0.1155}\\
			\hline
			1&(2,1)&0.4061 &0.4051 &0.4031 &\textbf{0.4007} &0.7102 &0.7083 &0.7049 &\textbf{0.7007} &0.1590 &0.1350 &0.1323 &\textbf{0.1169}\\
			2&(2,1)&0.4014 &0.4003 &0.3990 &\textbf{0.3980} &0.7018 &0.7000 &0.6976 &\textbf{0.6958} &\textbf{0.0762} &0.0838 &0.0843 &0.1688\\
			3&(2,1)&0.4008 &0.3994 &0.3985 &\textbf{0.3972} &0.7008 &0.6983 &0.6968 &\textbf{0.6945} &\textbf{0.0500} &0.0626 &0.0609 &0.0825\\
			\hline
			1&(2,3)&0.3602 &0.3595 &0.3594 &\textbf{0.3590} &0.6300 &0.6288 &0.6286 &\textbf{0.6279} &\textbf{0.2502} &0.3005 &0.3063 &0.2945\\
			2&(2,3)&0.3591 &0.3586 &0.3574 &\textbf{0.3547} &0.6279 &0.6269 &0.6248 &\textbf{0.6199} &\textbf{0.1470} &0.2033 &0.2045 &0.3189\\
			3&(2,3)&0.3589 &0.3581 &0.3571 &\textbf{0.3513} &0.6274 &0.6260 &0.6242 &\textbf{0.6139} &\textbf{0.0930} &0.1155 &0.1215 &0.2871\\
			\hline
			1&(2,4)&0.3510 &0.3508 &0.3505 &\textbf{0.3488} &0.6138 &0.6134 &0.6130 &\textbf{0.6100} &\textbf{0.2947} &0.3652 &0.3621 &0.3163\\
			2&(2,4)&0.3484 &0.3463 &0.3454 &\textbf{0.3433} &0.6091 &0.6055 &0.6039 &\textbf{0.6002} &\textbf{0.1770} &0.2768 &0.2967 &0.3277\\
			3&(2,4)&0.3469 &0.3449 &0.3433 &\textbf{0.3404} &0.6066 &0.6027 &0.5998 &\textbf{0.5947} &\textbf{0.1410} &0.1948 &0.1853 &0.2612\\
			\hline
			1&(3,4)&0.3497 &0.3488 &0.3486 &\textbf{0.3474} &0.6116 &0.6101 &0.6096 &\textbf{0.6075} &\textbf{0.3346} &0.3957 &0.3911 &0.3670\\
			2&(3,4)&0.3472 &0.3445 &0.3437 &\textbf{0.3411} &0.6069 &0.6023 &0.6008 &\textbf{0.5964} &\textbf{0.2051} &0.3004 &0.3063 &0.3796\\
			3&(3,4)&0.3456 &0.3436 &0.3420 &\textbf{0.3388} &0.6043 &0.6004 &0.5976 &\textbf{0.5919} &\textbf{0.1829} &0.2386 &0.2344 &0.3621\\
			\bottomrule[2pt]
		\end{tabular}
	}
\end{table}

\subsection{Macroeconomic indices of China}
In the second real example, we analyze a macroeconomic  dataset including 30 provinces of China. The dataset contains 81 macroeconomic indices across 30 provinces except Tibet, Taiwan, Hong Kong and Macao (which will be treated as zero in the following procedure) over 87 months from 2013-01 to 2023-03. The macroeconomic indices include consumer price index, finance, industry, trade, real estate and so on, see the supplement for further details. We imputed the missing values by the factor-model-based method, motivated by \cite{Yu2021Projected}. We applied difference operators to each  time series to guarantee stationarity. We further standardized the  dataset for further analysis. Figure \ref{fig:4} (b) also depicts the histogram of the sample kurtosis for this dataset, from which we can infer that the data are heavy-tailed and thus the robust methods may be more appropriate.

We first determine the number of factors. The IHR-RM suggests $(k_{1},k_{2})=(1,3)$, and all the other methods suggest $(k_{1},k_{2})=(1,1)$. For better illustration, we take $(k_{1},k_{2})=(2,3)$ in the following study. The estimated loading matrices after varimax rotation and scaling are plotted in Figure \ref{fig:5} and Figure  \ref{fig:6}. For the row factors, Figure \ref{fig:5} shows that they are closely related to the geographical location. The neighboring provinces tend to load similarly on the factors, such as Heilongjiang province, Jilin province and Liaoning province. For the column factors, it can be seen from Figure \ref{fig:6} that the macroeconomic indices can be divided into 4 groups.

To be consistent with Section 5.1, we also apply a rolling procedure to check the validation errors of different methods. For each month $t$ from 36 to 73, we repeatly use the $n \in \{1,2,3\}$ years observations and the numbers of factors $(k_{1},k_{2} \in \{1,2,3,4\})$ before $t$ to estimate the loading matrices and factors. And then we estimate the corresponding residuals of the following 12 months. Table \ref{table:6} (see further results  in the supplement) shows that IHR method preforms much more stably compared with the other methods in terms of the reconstruction errors and unexplained proportion of variances.

	\section{Discussion}
	
In this paper we focus on robust statistical inference for matrix factor model. An iterative Huber regression algorithm is proposed to estimate the row/column factor loadings and factor scores.
   We derive the convergence rates of the robust estimators for loadings, factors and common components  under finite second moment assumption of the idiosyncratic errors.   We also derive the asymptotic distributions of the estimators under mild conditions. To determine the pair of factor numbers, we propose a rank minimization and an eigenvalue-ratio method and the resultant estimators are proven to be consistent. 
   Numerical studies show that the proposed iterative Huber regression algorithm have great advantage over existing ones especially under the  heavy-tailed cases. An R package ``HDMFA" implementing the related robust matrix factor
analysis methods in the literature is available on CRAN. Theoretical property of the estimators
in the solution path of the iterative algorithm is still unknown and this problem is quite interesting but challenging, which we leave as a future research direction.

	\bibliographystyle{ref.bib}
	\bibliography{Ref}

\end{document}